# INACCURACY ASSESSMENT FOR SIMULTANEOUS MEASUREMENTS OF RESISTIVITY AND PERMITTIVITY APPLYING SENSITIVITY AND TRANSFER FUNCTION APPROACHES

A. Settimi, A. Zirizzotti, J. A. Baskaradas, C. Bianchi

INGV (Istituto Nazionale di Geofisica e Vulcanologia) –

via di Vigna Murata 605, I-00143 Rome, Italy

# Abstract

This paper proposes a theoretical modelling of the simultaneous and non invasive measurement of electrical resistivity and dielectric permittivity, using a quadrupole probe on a subjacent medium. A mathematical-physical model is applied on propagation of errors in the measurement of resistivity and permittivity based on the sensitivity functions tool. The findings are also compared to the results of the classical method of analysis in the frequency domain, which is useful for determining the behaviour of zero and pole frequencies in the linear time invariant (LTI) circuit of the quadrupole. The paper underlines that average values of electrical resistivity and dielectric permittivity may be used to estimate the complex impedance over various terrains and concretes, especially when they are characterized by low levels of water saturation (content) and analyzed within a bandwidth ranging only from low (LF) to middle (MF) frequencies. In order to meet the design specifications which ensure satisfactory performances of the probe (inaccuracy no more than 10%), the forecasts provided by the sensitivity functions approach are less stringent than those foreseen by the transfer functions method (in terms of both a larger band of frequency $f$ and a wider measurable range of resistivity $\rho$ or permittivity $\varepsilon_r$).

**Keywords**

Explorative geophysics;

Methods of non-destructive testing;

Complex impedance measurements: error theory.

# 1. Introductive review.

*Electrical resistivity survey in soil science* - The electrical resistivity of a surface is a proxy for the spatial and temporal variability of many other physical properties of the subjacent medium. Samouëlian (Samouëlian et al., 2005) discusses the basic principles of data interpretation and the main advantages or limits of the analysis. This method allows non-destructive and very sensitive investigation, describing subsurface properties without direct inspection. Various techniques are applied according to the required scales of resolution and heterogeneities of the area. A suitable probe injects generated electric currents into a medium and the resulting potential differences are measured. The information is recovered from the potential difference patterns, which provide the form of medium heterogeneities and their electrical properties (Kearey et al., 2002). The greater the electrical contrast between the subsurface matrix and a heterogeneity, the easier the detection. Other authors (Banton et al. 1997) showed that surface resistivity can be considered as a good indication of the variability of other physical properties. The current pattern distributions depend on the medium heterogeneities and are concentrated in a conductive volume. Some linear distributed arrays use four-electrode cells, which are commonly employed in the laboratory for resistivity calibration (Rhoades et al., 1976) and in the field for vertical electrical sounding (Loke, 2001).

*Dielectric permittivity survey in soil science [Middle frequencies (MF), 300kHz<f<3MHz]* - Analysis in middle frequencies allows the measurement of dielectric permittivity. Fechant and Tabbagh (Fechant and Tabbagh, 1999) developed an interesting approach. They used a MF band for the characterization of permittivity in the natural media. This approach employs an electrostatic quadrupole probe designed to measure resistivity at several centuries of *kHz* (Tabbagh, 1994). A quadrupole, working at the frequency *455kHz*, measures permittivity for determination of water content (Fechant, 1996). However, this approach requires calibration in laboratory.

*Electrical resistivity and dielectric permittivity surveys in soil science [low frequencies (LF), 30kHz<f<300kHz]* - Analysis in low frequencies allows simultaneous measurements of both

electrical resistivity and dielectric permittivity. Tabbagh and Grard, in their experiments (Grard, 1990, a-b)(Grard and Tabbagh, 1991)(Tabbagh et al., 1993), showed that the resistivity and dielectric constant (complex permittivity) of a surface can be measured by a set of four electrodes. This novel approach, first introduced by Wenner, improved the existing system which provided only a resistivity assessment. In the new method the four electrodes are manually inserted into the subjacent medium. Permittivity, which is sensitive to the presence of water, can also be determined employing a LF probe (below *300 kHz*) and plays an important role in the detection of anomalies in the subsurface.

Vannaroni and Del Vento (Vannaroni et al., 2004)(Del Vento and Vannaroni, 2005) used a dielectric spectroscopy probe to determine the complex permittivity of a surface from measurements of transfer impedance of a four-electrode system electrically coupled to the medium. They defined transfer impedance as the ratio of the voltage measured across a pair of receiving electrodes to the current transmitted by a second pair of electrodes (Vannaroni et al., 2004). This impedance measurement, performed in AC regime capacitive coupling, strongly depends on the geometry of the electrode array but also on the complex permittivity of the subsurface. The advantages offered by this method are due to the fact that the exciter current can be injected into the surface even in the absence of galvanic contact, and, in AC regime, both conduction and displacement currents of the medium can be measured, obtaining further information on the polarizability. In this case the frequency band is *10kHz-1MHz*. The lower limit is effectively imposed by two facts: a) firstly, the Maxwell-Wagner effect which limits probe accuracy (Frolich, 1990): the most important limitation happens because of interface polarization effects that are stronger at low frequencies, say below *1kHz* depending of medium conductivity; b) secondly, the need to maintain the amplitude of the current at measurable levels as, given the capacitive coupling between electrodes and soil, the current magnitude is proportional to the frequency. On the other hand, the upper limit is opportunely fixed to allow the analysis of the system in a regime of quasi static approximation and neglect the velocity factor of the cables used for the electrode harness, that in turn degrades the accuracy of the mutual impedance phase measurements. Thus, it is possible to exploit the analysis of the system in the low and middle frequency band where the electrostatic term results considerable. The general electromagnetic (e.m.) calculation provides lower values than the static case; a high resistivity narrows the differences. So, comparing, above *1 MHz* the general e.m. calculation must be preferred, while under *500 kHz* the static case should be used and between *500 kHz* and *1 MHz* both methods could be used (Tabbagh et al. 1993).

The present paper proposes a theoretical modelling of the simultaneous and non invasive measurement of electrical resistivity and dielectric permittivity, using a quadrupole probe on a subjacent medium. A mathematical-physical model is applied on propagation of errors in the measurement of resistivity and permittivity based on the sensitivity functions tool. The findings are also compared to the results of the classical method of analysis in the frequency domain, which is useful for determining the behaviour of zero and pole frequencies in the linear time invariant (LTI) circuit of the quadrupole. This paper underlines that average values of electrical resistivity and dielectric permittivity may be used to estimate the complex impedance over various terrains and concretes, especially when they are characterized by low levels of water saturation or content (Knight and Nur, 1987) and analyzed within a frequency bandwidth ranging only from LF to MF (Myounghak et al., 2007)(Al-Qadi et al., 1995). In order to meet the design specifications which ensure satisfactory performances of the probe (inaccuracy no more than *10%*), the forecasts provided by the theory of error propagation [suggested by (Vannaroni et al., 2004)] applying the sensitivity functions approach, explicitly developed in the paper, are less stringent than those foreseen by the analysis in the frequency domain [suggested by (Grard and Tabbagh, 1991)], deepening here the transfer function method to analyze the zero and pole behaviour (in terms of both a larger band of frequency $f$ and a wider measurable range of resistivity $\rho$ or permittivity $\varepsilon_r$).

The paper is organized as follows. Section 2 discusses the Cole-Cole empiric function: for simplicity of analysis, the dielectric dispersion is assumed very low: this operating condition is satisfied when the electrical spectroscopy is performed only on non-saturated water materials and especially in a suitable band of low and middle frequencies. Section 3 introduces the quadrupole probe. Sec. 4 provides a theoretical modelling which applies to both the sensitivity functions approach (Sec 4.a.) and the transfer function method (4.b.). In Sec 5, the configurations of the quadrupole are defined and discussed. Conclusions are drawn in Sec. 6. Finally, an outline of the somewhat lengthy calculations is presented in the Appendices A and B.

## 2. Discussing Cole-Cole empiric function.

Even if, according to Debye polarization mechanisms (Debye, 1929) or Cole-Cole diagrams (Auty and Cole, 1952), the complex permittivity of various materials in the frequency band from VLF to VHF exhibits several intensive relaxation effects and a non-trivial dependence on the water saturation (Chelidze and Gueguen, 1999)(Chelidze et al., 1999), anyway average values of electrical resistivity and dielectric permittivity may be used to estimate the complex impedance over various terrains and concretes, especially when they are characterized by low levels of water content (Knight and Nur, 1987) and analyzed within a frequency bandwidth ranging only from LF to MF (Myounghak et al., 2007)(Al-Qadi et al., 1995).

Many functions have been proposed to fit the data of dielectrics. Among them, there are those obtained by attempts to model the physical processes or those of simple empirical functions, which are used to parameterize the data without the knowledge of the involved mechanisms. A widely used empirical function has been proposed by brothers Cole and it is based on the theory of Debye relaxation, the first one to have treated this phenomenon.

The Cole-Cole empiric function defines the first order dielectric response of materials in frequency domain,

$$\varepsilon_r^{complex}(f) = \varepsilon_r(f) - j\frac{\sigma(f)}{2\pi f \varepsilon_0} = \varepsilon_{r,H} + \frac{\varepsilon_{r,L} - \varepsilon_{r,H}}{1+(j2\pi f \tau)^{1-\alpha}} - j\frac{\sigma_L}{2\pi f \varepsilon_0}, \qquad (2.1)$$

consisting of real and imaginary parts:

$$\frac{\varepsilon_r(f) - \varepsilon_{r,H}}{\varepsilon_{r,L} - \varepsilon_{r,H}} = \frac{1 + (2\pi f \tau)^{1-\alpha} \sin(\alpha \pi/2)}{1 + (2\pi f \tau)^{2(1-\alpha)} + 2(2\pi f \tau)^{1-\alpha} \sin(\alpha \pi/2)}, \tag{2.2}$$

$$\frac{[\sigma(f) - \sigma_L]/2\pi f \varepsilon_0}{\varepsilon_{r,L} - \varepsilon_{r,H}} = \frac{(2\pi f \tau)^{1-\alpha} \cos(\alpha \pi/2)}{1 + (2\pi f \tau)^{2(1-\alpha)} + 2(2\pi f \tau)^{1-\alpha} \sin(\alpha \pi/2)}, \tag{2.3}$$

where $\varepsilon_0$ is the dielectric constant in vacuum.

The electrical conductivity $\sigma(f)$ and dielectric permittivity $\varepsilon_r(f)$ exhibit limit values at low and high frequencies, $\sigma_L$, $\varepsilon_{r,L}$ and $\sigma_H$, $\varepsilon_{r,H}$, which are linked by the relaxation time $\tau$,

$$\frac{(\varepsilon_L - \varepsilon_H)\varepsilon_0}{\tau} = \sigma_H - \sigma_L, \tag{2.4}$$

so, as it can be noticed, permittivity and conductivity can not vary independently of each other (Frolich, 1990).

At the characteristic frequency of relaxation, $f_c = 1/(2\pi \cdot \tau)$, the permittivity $\varepsilon_r$ assumes an intermediate value between the values of high and low frequency, $\varepsilon_{r,L}$ and $\varepsilon_{r,H}$. Alternately, the relaxation frequency $f_c$ could be considered as that frequency at which the conductivity $\sigma$ assumes the middle value between the two limit values, $\sigma_L$ and $\sigma_H$.

In reality, the relationship (2.1) is a generalization of Debye equation, having the purpose to take into account, through the introduction of another parameter α (inclusive between 0 e 1), the enlargement of dispersion region due to the complexity of structure and the composition of materials. Note that, for $\alpha = 0$, eq. (2.1) can be exactly reduced to Debye equation. It is to be underlined that the parameter α is a increasing function of the water saturation $S_W$, such that $\alpha(S_W=0) \to 0$, reaching a limit value $\alpha_L > 0$ for $S_W \to 1$ (Knight and Nur, 1987). In fact, the complex dielectric permittivity is flattened decreasing water content or increasing frequency (Myounghak et al., 2007)(Al-Qadi et al., 1995).

The complex dielectric permittivity $\varepsilon_r^{complex}(f)$ can be approximated to a constant if its dominant term $(\varepsilon_{r,L} - \varepsilon_{r,H})/[1 + (j2\pi f \tau)^{1-\alpha}]$ is a function almost independent from the frequency,

$$(2\pi f \tau)^{1-\alpha} \ll 1. \tag{2.5}$$

Operating condition (2.5) holds when the materials are characterized by low levels of water content, i.e.

$$\alpha \to 0, \qquad (2.6)$$

and analyzed over a band lower than middle frequencies, i.e.

$$f < \frac{1}{2\pi\tau}. \qquad (2.7)$$

In fact, the constant $\tau$ depends on the physical process under consideration and it has an order of magnitude that varies from a few picoseconds for the orientation of electrons and small dipolar molecules, up to a few seconds for the effects of counter-ions or for the interfacial polarization (Frolich, 1990).

So, in the paper, let us refer to the *(σ, ε$_r$)* values in LF-MF bandwidth proposed for various terrains in (Edwards, 1998) and for concretes in (Polder et al., 2000)(Laurents, 2005).

## 3. Quadrupole probe.

When using a quadrupole probe [fig. 1] the response depends on geometric parameters, like the height of each electrode above the ground surface and the separation of the electrodes, and on physical parameters including frequency, electrical conductivity and dielectric permittivity. When a medium is assumed to be linear and its response linearly dependent on the electrical charges of the two exciting electrodes, the simplest approach is static calculation (Tabbagh et al., 1993), especially using a low operating frequency. If the electrodes have small dimensions relative to their separations, then they can be considered as points. Moreover, if the current wavelength is much larger than all the dimensions under consideration, then quasi-static approximation applies (Grard, 1990, a-b).

The quadrupole probe [fig. 1] measures a capacitance in vacuum $C_0(L)$, which is directly proportional to its characteristic geometrical dimension, i.e. the electrode-electrode distance $L$, both in a linear (Wenner) configuration [fig. 2.a],

$$C_0(L) = 4\pi\varepsilon_0 \cdot L, \qquad (3.1)$$

and in a square arrangement [fig. 2.b],

$$C_0(L) = \alpha \cdot 4\pi\varepsilon_0 L, \qquad (3.2)$$

which is greater by a factor $\alpha=1/(2-2^{1/2})>1$, where $\varepsilon_0$ is the dielectric constant in vacuum. When the quadrupole, specified by the electrode-electrode distance $L$, has a galvanic contact with the subjacent medium, of electrical conductivity $\sigma$ and dielectric permittivity $\varepsilon_r$, it measures a transfer impedance $Z_N(f,L,\sigma,\varepsilon_r)$, which consists of parallel components of resistance $R_N(L,\sigma)$ and capacitance $C_N(L,\varepsilon_r)$. The resistance $R_N(L,\sigma)$ depends only on $L$ and $\sigma$ (Grard and Tabbagh, 1991)

$$R_N(L,\sigma) = 2\frac{\varepsilon_0/\sigma}{C_0(L)}; \qquad (3.3)$$

while $C_N(L,\varepsilon_r)$ depends only on $L$ and $\varepsilon_r$ (Grard and Tabbagh, 1991)

$$C_N(L,\varepsilon_r) = \frac{1}{2}C_0(L) \cdot (\varepsilon_r + 1). \qquad (3.4)$$

As a consequence, if the probe, besides grazing the medium, measures the conductivity $\sigma$ and permittivity $\varepsilon_r$ working in a frequency $f$ much lower than the cut-off frequency $f_T=f_T(\sigma,\varepsilon_r)= \sigma/(2\pi\varepsilon_0(\varepsilon_r+1))$, the transfer impedance $Z_N(f,L,\sigma,\varepsilon_r)$ is characterized by the phase $\Phi_N(f,\sigma,\varepsilon_r)$ and modulus $|Z|_N(L,\sigma)$. The phase $\Phi_N(f,\sigma,\varepsilon_r)$ depends linearly on $f$ with a maximum value of $\pi/4$ and is directly proportional to the ratio $(\varepsilon_r+1)/\sigma$; while $|Z|_N(L,\sigma)$ does not depends on $f$, and is inversely proportional to both $L$ and $\sigma$. In fact, if $Z_N(f,L,\sigma,\varepsilon_r)$ consists of the parallel components of $R_N(L,\sigma)$ (3.3) and $C_N(L,\varepsilon_r)$ (3.4), then it is fully characterized by the high frequency pole $f_T=f_T(\sigma,\varepsilon_r)$, which cancels its denominator: the transfer impedance acts as a low-middle frequency band-pass filter with cut-off $f_T=f_T(\sigma,\varepsilon_r)$, in other words the frequency equalizing Joule and displacement current. In the operating conditions defined in sec 2, average values of $\sigma$ may be used over the band ranging from LF to MF, therefore $|Z|_N(L,\sigma)$ is not function of frequency below $f_T$.

Instead, when the quadrupole probe [fig. 1] has a capacitive contact with the subjacent medium and the geometry of the probe is characterized by the ratio $x$ between the height above ground $h$ and the electrode-electrode distance $L$,

$$x = \frac{h}{L}, \qquad (3.5)$$

its configurations can be entirely defined by a suitable geometrical factor $K(x)$, which depends on the height/dimension ratio $x$. It is introduced by (Grard and Tabbagh, 1991) and can be specified for the Wenner configuration [fig. 2.a]

$$K(x) = 2(1+4x^2)^{-1/2} - (1+x^2)^{-1/2}, \qquad (3.6)$$

and the square arrangement [fig. 2.b]

$$K(x) = \frac{(1+4x^2)^{-1/2} - 2^{-1/2}(1+2x^2)^{-1/2}}{1-2^{-1/2}}; \qquad (3.7)$$

Actually, Grard and Tabbagh preferred to introduce the complementary $\delta(x)$ of the geometrical factor $K(x)$, i.e.

$$\delta(x) = 1 - K(x), \qquad (3.8)$$

where $K(x=0)=1$ and $\delta(x=0)=0$.

So, if the quadrupole works in the pulse frequency $\omega=2\pi f$, which can be normalized with respect to the cut-off $\omega_T=2\pi f_T$ (Grard and Tabbagh, 1991),

$$\Omega = \frac{\omega}{\omega_T} = \omega R_N C_N = \omega \frac{\varepsilon_0(\varepsilon_r+1)}{\sigma}, \qquad (3.9)$$

then the probe measures a transfer impedance $Z(\Omega,x,\sigma,\varepsilon_r)$ which consists of the resistance $R(\Omega,x,\sigma,\varepsilon_r)$ and capacitance $C(\Omega,x,\sigma,\varepsilon_r)$ parallel components (Grard and Tabbagh, 1991),

$$R(\Omega,x,\sigma,\varepsilon_r) = R_N(L,\sigma) \frac{[1+\delta(x)\frac{\varepsilon_r-1}{2}]^2 + [\frac{\delta(x)}{\Omega}\frac{\varepsilon_r+1}{2}]^2}{1-\delta(x)}, \qquad (3.10)$$

$$C(\Omega, x, \sigma, \varepsilon_r) = C_N(L, \varepsilon_r) \frac{1 + \delta(x)(\frac{\varepsilon_r - 1}{2} + \frac{1}{\Omega^2}\frac{\varepsilon_r + 1}{2})}{[1 + \delta(x)\frac{\varepsilon_r - 1}{2}]^2 + [\frac{\delta(x)}{\Omega}\frac{\varepsilon_r + 1}{2}]^2}. \qquad (3.11)$$

Inverting eqs. (3.10) and (3.11), $\sigma$ and $\varepsilon_r$ can be expressed as functions of $R$ and $C$, i.e.

$$\sigma(\omega, x, R, C) = \frac{2[1 - \delta(x)]\varepsilon_0 \omega^2 R C_0}{\delta^2(x) + \omega^2 R^2 [C_0 - \delta(x)C]^2}, \qquad (3.12)$$

$$\varepsilon_r(\omega, x, R, C) = \frac{\delta(x)[\delta(x) - 2] - \omega^2 R^2 [C_0 - \delta(x)C]\{C_0 + [\delta(x) - 2]C\}}{\delta^2(x) + \omega^2 R^2 [C_0 - \delta(x)C]^2}. \qquad (3.13)$$

In our opinion, once fixed the pair's *(f, x)* degrees of freedom, it is not suitable to choose *(R,C)* as independent variables and then *(σ, ε_r)* as dependent variables [eqs. (3.12)-(3.13)]. Instead, it is more appropriate to consider *(σ, ε_r)* as quantities of physical interest and consequently eqs. (3.10)-(3.11) as starting point for the theoretical development. In fact, even if the physics does not forbid to choose *(R,C)* as independent variables, running the way *(R,C) → (σ, ε_r)*, anyway the procedures of design should choose *(σ, ε_r)* as independent variables, running a preferential way *(σ, ε_r) → (R,C)*. According to the two following practical approaches: a) [*(σ, ε_r)* as independent variables in order] to establish the class of media with conductivity and permittivity *(σ, ε_r)* which are investigable by a quadrupole working in a fixed band *B* and specified by a known geometry *x*; b) [preferential way *(σ, ε_r) → (R,C)* since] once a subjacent medium with electrical conductivity *σ* and dielectric permittivity *ε_r* is selected, one can project the quadrupole probe specifications *R* and *C* both in frequency *f* and in height/dimension ratio *x*.

## 4. Theoretical modelling.

The measurements taken using the quadrupole probe are affected by errors mainly originating from uncertainties associated with transfer impedance, from dishomogeneities between the modelled and actual stratigraphy, and from inaccuracy of the electrode array deployment above the surface (Vannaroni et al., 2004). Errors in impedance result mainly from uncertainties in the electronic systems that perform the amplitude and phase measurements of the voltages and currents (Del Vento and Vannaroni, 2005). The above uncertainties were assumed constant throughout the whole frequency band even though their effects, propagating through the transfer function, will produce a frequency dependent perturbation.

*4.1. Sensitivity Functions Approach .*

This paper proposes to develop explicitly the sensitivity function approach which is implied in the theory of error propagation suggested by (Vannaroni et al., 2004). In fact, the section introduces a mathematical-physical model for the propagation of errors in the measurement of electrical conductivity $\sigma$ and dielectric permittivity $\varepsilon_r$, based on the sensitivity functions tool (Murray-Smith, 1987). This is useful for expressing inaccuracies in the measurement of conductivity and permittivity [fig. 3] as a linear combination of the inaccuracies for the transfer impedance, both in modulus $|Z|$ and in phase $\Phi_Z$, where the weight functions are inversely proportional just to the sensitivity functions for $|Z|$ and $\Phi_Z$ relative to $\sigma$ and $\varepsilon_r$ [fig. 4]. The inaccuracies of transfer impedance depend on the inaccuracies of electrical voltage and current which are assigned by the employed electronics and, in particular, by the sampling methods.

So, the inaccuracies $\Delta\sigma/\sigma$, in the measurement of the electrical conductivity $\sigma$, and $\Delta\varepsilon_r/\varepsilon_r$, in the dielectric permittivity $\varepsilon_r$, can be expressed as a linear combination of the inaccuracies $\Delta|Z|/|Z|$ and $\Delta\Phi_Z/\Phi_Z$ in the measurement of the transfer impedance, respectively in modulus $|Z|$ and in phase $\Phi_Z$,

$$\frac{\Delta\sigma}{\sigma} = \left|S_{|Z|}^{\sigma}\right|\frac{\Delta|Z|}{|Z|} + \left|S_{\Phi_Z}^{\sigma}\right|\frac{\Delta\Phi_Z}{\Phi_Z} = \frac{1}{\left|S_{\sigma}^{|Z|}\right|}\frac{\Delta|Z|}{|Z|} + \frac{1}{\left|S_{\sigma}^{\Phi_Z}\right|}\frac{\Delta\Phi_Z}{\Phi_Z} \quad , \quad \text{for} \quad \varepsilon_r = const, \qquad (4.1)$$

$$\frac{\Delta\varepsilon_r}{\varepsilon_r} = \left|S_{|Z|}^{\varepsilon_r}\right|\frac{\Delta|Z|}{|Z|} + \left|S_{\Phi_Z}^{\varepsilon_r}\right|\frac{\Delta\Phi_Z}{\Phi_Z} = \frac{1}{\left|S_{\varepsilon_r}^{|Z|}\right|}\frac{\Delta|Z|}{|Z|} + \frac{1}{\left|S_{\varepsilon_r}^{\Phi_Z}\right|}\frac{\Delta\Phi_Z}{\Phi_Z} \quad , \quad \text{for} \quad \sigma = const, \qquad (4.2)$$

where ($S_\sigma^{|Z|}, S_\sigma^{\Phi_Z}$) and ($S_{\varepsilon_r}^{|Z|}, S_{\varepsilon_r}^{\Phi_Z}$) are the pairs of sensitivity functions for the transfer impedance, both in |Z| and $\Phi_Z$, relative to the conductivity $\sigma$ and permittivity $\varepsilon_r$, whose expressions are reported in Appendix A. The conditions $\sigma=const$ and $\varepsilon_r=const$ in eqs. (4.1) and (4.2) underline not so much that constant values of electrical conductivity and dielectric permittivity are used to estimate the complex impedance over various terrains and concretes under the operating conditions defined in Sec. 2, as that the quantities $\sigma$ and $\varepsilon_r$ are not independent of each other, since the electrical displacement shows a phase-shift with respect to the electrical field (Frolich, 1990); so, for need to distinguish the inaccuracies in measurements of conductivity and permittivity, the inaccuracy $\Delta\sigma/\sigma$ can only be calculated assuming there is not uncertainty for $\varepsilon_r$ ($\Delta\varepsilon_r=0 \Leftrightarrow \varepsilon_r=const$) and vice versa. Moreover, according to the physical problem, the probe performs measurements of the transfer impedance $Z$, both in modulus |Z| and in phase $\Phi_Z$, which are characterized by the inaccuracies $\Delta|Z|/|Z|>0$ and $\Delta\Phi_Z/\Phi_Z>0$. Mathematically, it is not allowed to apply the conditions |Z|=const or $\Phi_Z=const$. In this context, the sensitivity functions $S_{|Z|}^\sigma$ and $S_{|Z|}^{\varepsilon_r}$ can not be calculated assuming $\Phi_Z=const$ and then the sensitivities $S_{\Phi_Z}^\sigma$ and $S_{\Phi_Z}^{\varepsilon_r}$ assuming |Z|=const. In fact, as discussed above, once fixed the pair's (f, x) degrees of freedom, it is not suitable to choose as independent variables (|Z|,$\Phi_Z$) [or (R,C)]. Consequently, the sensitivity functions can not be calculated by the dependent variables $\sigma=\sigma(|Z|, \Phi_Z)$ and $\varepsilon_r=\varepsilon_r(|Z|, \Phi_Z)$ [or by eqs. (3.12) and (3.13)]. Instead, the physical problem should be approached recalling that (f, x, $\sigma$, $\varepsilon_r$) have been considered as independent variables. In the simplifying hypothesis that the frequency f and the height/dimension ratio x are characterized by inaccuracies $\Delta f/f \approx 0$ and $\Delta x/x \approx 0$ close to zero, the conditions f=const and x=const can be applied. Necessarily, the inaccuracy $\Delta\sigma/\sigma$, in the measurement of the electrical conductivity $\sigma$, is calculated assuming $\varepsilon_r=const$, and then the inaccuracy $\Delta\varepsilon_r/\varepsilon_r$ for the dielectric permittivity $\varepsilon_r$, assuming $\sigma=const$. As a consequence, the mathematical calculations should be done recalling the fact that eqs. (3.10)-(3.11) have been considered as starting point for the theoretical development. The inaccuracies $\Delta\sigma/\sigma$ for the conductivity $\sigma$ and $\Delta\varepsilon_r/\varepsilon_r$ for the permittivity $\varepsilon_r$ can be more directly expressed as functions of *(f, x, σ, ε_r)* by calculating the sensitivity functions ($S_\sigma^{|Z|}, S_\sigma^{\Phi_z}$) and ($S_{\varepsilon_r}^{|Z|}, S_{\varepsilon_r}^{\Phi_z}$) in the last member of eqs. (4.1) and (4.2). These sensitivities are derived from the transfer impedance $1/Z=1/R+j\omega C$ reported in eqs. (3.10) and (3.11).

The interesting physical results obtained using this sensitivity functions approach are discussed below. If the quadrupole probe has a galvanic contact with the subjacent medium, i.e. *h=0*, then the inaccuracies *Δσ/σ* in the measurement of the electrical conductivity *σ* and *Δε_r/ε_r* in the dielectric permittivity *ε_r* are minimized in the frequency band *B* of the quadrupole, for all its geometric configurations and media; and, even if *h≠0*, the design of the probe must still be optimized with respect to the minimum value of the inaccuracy *Δε_r/ε_r* in *ε_r*, which is always higher than the corresponding minimum value of the inaccuracy *Δσ/σ* in the band *B* of the probe, for all its configurations and media (Tabbagh et al., 1993)(Vannaroni et al., 2004).

Under quasi static approximation, only if the quadrupole probe is in galvanic contact with the subjacent medium, i.e. *h=0*, and considering that the sensitivities functions are defined as normalized functions, then our mathematical-physical model predicts that the sensitivities of the transfer impedance relative to the conductivity *σ* and permittivity *ε_r* are independent of the characteristic geometrical dimension of the quadrupole, i.e. electrode-electrode distance *L*.

If the probe grazes the medium, then the transfer impedance $Z_N(\sigma,L)$ consists of the resistance $R_N(\sigma,L)$, which is independent of *ε_r*, and parallel capacitance $C_N(\varepsilon_r,L)$, which is independent of *σ*, such that: the sensitivity function $S_\sigma^R$ for *R* relative to *σ* is a constant equal to *(-1)*; the sensitivity $S_{\varepsilon_r}^C(\varepsilon_r)$ for *C* relative to *ε_r* is independent of *σ*, behaving as the function *ε_r/(ε_r+1)* of *ε_r*; the $S_{\varepsilon_r}^R$ function for *R* relative to *ε_r* and the $S_\sigma^C$ function for *C* relative to *σ* are identically null. As a consequence, the inaccuracy *ΔR/R* for *R* shows the same behaviour versus frequency of the inaccuracy *Δσ/σ* in the measurement of *σ*, as *ΔR/R=|$S_\sigma^R$|Δσ/σ=Δσ/σ*, and the inaccuracy *ΔC/C* for *C* shows a similar behaviour versus frequency with respect to the inaccuracy *Δε_r/ε_r* for *ε_r*, as

$\Delta C/C = |S^C_{\varepsilon_r}(\varepsilon_r)| \Delta\varepsilon_r/\varepsilon_r \approx \Delta\varepsilon_r/\varepsilon_r$ if $\varepsilon_r \gg 1$. Moreover, besides the hypothesis $h=0$, if $\sigma$ and $\varepsilon_r$ are measured in the cut-off frequency $f_T=f_T(\sigma,\varepsilon_r)$, then: the sensitivity functions $S^{|Z|}_\sigma$ and $S^{\Phi_Z}_\sigma$ for the transfer impedance, both in modulus $|Z|$ and in phase $\Phi_Z$, relative to $\sigma$, are constant, respectively $(-1/4)$ and $(-1/\pi)$; the sensitivities $S^{|Z|}_{\varepsilon_r}(\varepsilon_r)$ and $S^{\Phi_Z}_{\varepsilon_r}(\varepsilon_r)$ for $|Z|$ and $\Phi_Z$ relative to $\varepsilon_r$ are independent of $\sigma$, such that they behave as the function $\varepsilon_r/(\varepsilon_r+1)$ of $\varepsilon_r$. As a consequence, the ratio between $\Delta\varepsilon_r/\varepsilon_r$ and $\Delta\sigma/\sigma$ is independent of $\sigma$, behaving as the function $(1+1/\varepsilon_r)$ of $\varepsilon_r$, and $\Delta\sigma/\sigma$ is a constant equal to $\Delta\sigma/\sigma = 4\Delta|Z|/|Z| + \pi\Delta\Phi_Z/\Phi_Z$. As post-test, only assuming the conditions $\sigma=const$ and $\varepsilon_r=const$ in eqs. (4.1) and (4.2), the sensitivity function approach provides results according to ref. (Vannaroni et al., 2004).

*4.2. Transfer Function method.*

This paper proposes to deepen the transfer function method, by analyzing the zero and pole behaviour, which is implied in the frequency domain analysis suggested by (Grard and Tabbagh, 1991). In fact, the section introduces the method of analysis in the frequency domain for determining the behaviour of the zero and pole frequencies in the LTI circuit of the quadrupole probe [fig. 1]. In order to satisfy the operative conditions of linearity for the measurements, if the quadrupole has a capacitive contact with the subjacent medium, then one should impose the frequency $f$ of the probe to be included between the zero $z_M$ and the pole $p_M$ of the transfer impedance, so its modulus to be almost constant within the frequency band (Grard and Tabbagh, 1991),

$$z_M(x,\varepsilon_r,\sigma) \leq f \leq p_M(x,\varepsilon_r,\sigma). \qquad (4.3)$$

Based on the above conditions, an optimization equation is deduced for the probe, which links the optimal ratio $x$ between its height above ground and its characteristic geometrical dimension only to the dielectric permittivity $\varepsilon_r$ of the medium, so that

$$\delta(x) \approx \frac{2}{15\varepsilon_r + 17}. \tag{4.4}$$

In order to satisfy the operative conditions of linearity for the measurements, if the quadrupole is in galvanic contact with the subjacent medium, then one should impose the working frequency $f$ of the quadrupole to be lower than the cut-off frequency of the transfer impedance, so its modulus to be constant below the cut-off frequency. Just under the above conditions, it is optimal to design the characteristic geometrical dimensions of the probe or establish the measurable ranges of the conductivity $\sigma$ and permittivity $\varepsilon_r$ of the medium [fig. 5]. The results (4.3) and (4.4), derived by the classical transfer function method, are demonstrated in the Appendix B.

The interesting physical results obtained using this transfer function method are discussed below. In order to meet the design specifications which ensure satisfactory performances of the probe (inaccuracy no more than *10%*), the forecasts provided by the theory of error propagation [suggested by (Vannaroni et al., 2004)] applying the sensitivity functions approach, explicitly developed in the paper, are less stringent than those foreseen by the analysis in the frequency domain [suggested by (Grard and Tabbagh, 1991)], deepening here the transfer function method to analyze the zero and pole behaviour (in terms of both a larger band of frequency $f$ and a wider measurable range of resistivity $\rho$ or permittivity $\varepsilon_r$) [figs. 6,7].

In fact, given a surface (for example, a non-saturated concrete with low conductivity $\sigma=10^{-4}$ *S/m* and $\varepsilon_r=4$) with dielectric permittivity $\varepsilon_r$ [fig. 6]:

- if the quadrupole probe has a capacitive contact with the subjacent medium, i.e. $h\neq0$, then, having defined an optimal ratio $x_{opt}=h_{opt}/L$ between an optimal height $h_{opt}$ above ground and the characteristic geometrical dimension $L$, the transfer impedance $Z(f,x_{opt})$ in units of $1/h_{opt}$, calculated in $x_{opt}$, is a function of the working frequency $f$ such that its modulus $|Z|(f,x_{opt})$, in units of $1/h_{opt}$, is almost constant between a zero frequency $z(x_{opt})$, almost one decade higher than a minimum frequency value $f_{min}(x_{opt})$ allowing the inaccuracy $\Delta\varepsilon_r/\varepsilon_r(f,x_{opt})$ in the measurement of $\varepsilon_r$ below a prefixed limit (*10%*), and a pole $p(x_{opt})$, almost one decade lower than the maximum value of frequency $f_{max}(x_{opt})$ satisfying the requirement that the inaccuracy $\Delta\varepsilon_r/\varepsilon_r(f,x_{opt})$ for $\varepsilon_r$ is below *10%* [fig. 6][fig. 8];

- if $h=0$, i.e. the quadrupole of electrode-electrode distance $L$ grazes a medium of conductivity $\sigma$, then the transfer impedance $Z(f,L)$, calculated in $L$, is a function of the working frequency $f$ such that its modulus $|Z|(f,L)$ is constant down to the cut-off frequency $f_T=f_T(\sigma,\varepsilon_r)$, which is higher than an optimal frequency $f_{opt}(L)$ minimizing the inaccuracy $\Delta\varepsilon_r/\varepsilon_r(f,L)$. Materials characterized by a low $\sigma$ or a high $\varepsilon_r$ lead to the effect of leftward shifting of the cut-off frequency $f_T$, so reducing the optimal frequency $f_{opt}(L)$ [fig. 9];

- usually, on a selected surface, it is possible to verify that the probe in capacitive contact performs optimal measurements over the band $[f_{min}(x_{opt})<z(x_{opt}), f_{max}(x_{opt})>p(x_{opt})]$, which is shifted towards lower and higher frequencies compared to the case when the probe is in galvanic contact, where the respective band $[f_{min}, f_{max}]$ is narrower of almost one decade in frequency, especially increasing the value of $\varepsilon_r$ [figs. 8.e, 9.c].

Moreover, once the frequency band $B$ is fixed [fig. 7]:

- if the quadrupole probe has a capacitive contact with the subjacent medium, then the ratio $x=h/L$, between the height $h$ above ground and the characteristic geometrical dimension $L$, ranges from the lower limit $x_{low}$, corresponding to water ($\varepsilon_r=81$).

  In a preliminary analysis, based on the transfer functions approach, it follows that the quadrupole, designed with the height/dimension ratio $x=h/L$, optimally measures dielectric permittivity $\varepsilon_{r,opt}$; the modulus $|Z|(x,\sigma,\varepsilon_{r,opt})$, in units of $1/h$, of its transfer impedance, calculated in $x$ and $\varepsilon_{r,opt}$, function of the electrical conductivity $\sigma$, is characterized by a zero $z(\sigma,\varepsilon_{r,opt})$ and a pole $p(\sigma,\varepsilon_{r,opt})$ frequency, which respectively fall near the lower and upper limit of $B$ when $\sigma$ is measured within a range of lower limit $\sigma'_{low}$ and upper limit $\sigma'_{up}$.

  In a deeper analysis, based on the sensitivity function method, it is possible to verify, still designing the quadrupole with the ratio $x=h/L$ for an optimal measurement of $\varepsilon_{r,opt}$, the measurable range of $\sigma$; the inaccuracy $\Delta\varepsilon_r/\varepsilon_r(x,\sigma,\varepsilon_{r,opt})$ in the measurement of $\varepsilon_{r,opt}$, a function of $\sigma$, is below a prefixed limit (*10%*) if $\sigma$ is measured within the range *[$\sigma_{low}$, $\sigma_{up}$]* larger than *[$\sigma'_{low}$, $\sigma'_{up}$]* by almost one magnitude order (both right and left side) [fig. 7] [fig. 10] [tabs. 1, 2].

- If *h=0*, i.e. the probe of electrode-electrode distance *L* grazes a medium of conductivity $\sigma$ and permittivity $\varepsilon_r$, then the transfer impedance *Z(L,$\sigma$,$\varepsilon_r$)*, calculated in *L*, is a function of $\sigma$ and $\varepsilon_r$ such that its cut-off frequency $f_T=f_T(\sigma,\varepsilon_r)$, a function of both $\sigma$ and $\varepsilon_r$, ranges from $f_{T,min}=100kHz$ to $f_{T,max}=1MHz$ for the materials belonging to an *($\sigma$,$\varepsilon_r$)*-domain, almost super-imposable with the corresponding one within which the inaccuracy $\Delta\varepsilon_r/\varepsilon_r(L,\sigma,\varepsilon_r)$ for $\varepsilon_r$ is below about *10%* [fig. 11].

- Usually, having fixed the frequency band, the probe in capacitive contact performs optimal measurements over surfaces of lower conductivities compared to the case when the probe is in galvanic contact, as the respective conductivities are higher even of almost one magnitude order [tabs. 1, 3].

## 5. Quadrupole configurations.

The transfer impedance of a quadrupolar array can be evaluated for any arbitrary configuration. As a general rule it is assumed that subsurface electrical sounding becomes scarcely effective at depths greater than the horizontal distance between the electrodes (Grard and Tabbagh, 1991)(Vannaroni et al., 2004). This paper considers two kinds of probes, i.e. square and linear (Wenner) configurations. The square configuration is an array of two horizontal parallel dipoles with the four electrodes positioned at the corners of a square (Grard and Tabbagh, 1991). Instead, the Wenner arrangement consists of four terminals equally spaced from one another along a straight horizontal line (Vannaroni et al., 2004).

If the quadrupole probe [fig. 1] is characterized by a characteristic geometrical dimension $L$, then the linear (Wenner) configuration [fig. 2.a] measures a capacitance in vacuum $C_{0,W}=4\pi\varepsilon_0\cdot L$, while in the square arrangement [fig. 2.b] $C_{0,S}=\alpha\cdot C_{0,W}$, greater by a factor $\alpha=1/(2-2^{1/2})>1$.

When the quadrupole is in galvanic contact, i.e. $h=0$, with a subjacent medium of electrical conductivity $\sigma$ and dielectric permittivity $\varepsilon_r$, the Wenner configuration measures a resistance $R_{N,W}=2\varepsilon_0/\sigma C_{0,W}$ and a parallel capacitance $C_{N,W}=C_{0,W}\cdot(\varepsilon_r+1)/2$, while in the square arrangement $R_{N,S}=R_{N,W}/\alpha$ and $C_{N,S}=\alpha\cdot C_{N,W}$, so, at the frequency $f$, the transfer impedance $1/Z_N=1/R_N+j2\pi f\cdot C_N$ for the Wenner configuration is defined by a modulus $|Z|_{N,W}=1/[(1/R_{N,W})^2+(2\pi f\cdot C_{N,W})^2]^{1/2}$ and a phase $\Phi_{N,W}=arctg(2\pi f\cdot R_{N,W}\cdot C_{N,W})$, while in the square arrangement $|Z|_{N,S}=Z|_{N,W}/\alpha$, smaller by a factor $1/\alpha$ [fig. 9.a] and $\Phi_{N,S}=\Phi_{N,W}$, which is maintained invariant in the Wenner or square configurations [fig. 9.b]. Also the cut-off frequency is independent of the configurations, i.e. $f_T=f_T(\sigma,\varepsilon_r)$. Moreover, if the probe grazes the medium and considering that the sensitivity functions are defined as normalized functions, then the sensitivities $S_\sigma^{|Z|}$ and $S_\sigma^{\Phi_z}$, relative to the conductivity $\sigma$, and the functions $S_{\varepsilon_r}^{|Z|}(\varepsilon_r)$ and $S_{\varepsilon_r}^{\Phi_z}(\varepsilon_r)$, relative to the permittivity $\varepsilon_r$, for the transfer impedance, both in modulus $|Z|$ and in phase $\Phi_Z$, are invariant in the Wenner or square configurations. Only if $h=0$, are the inaccuracies $\Delta\sigma/\sigma$ in the measurement of $\sigma$ and $\Delta\varepsilon_r/\varepsilon_r$ for $\varepsilon_r$ also independent of the configurations, so the probe is characterized by the same performances in the frequency band $B$ and in the measurable ranges of $\sigma$ and $\varepsilon_r$ [fig. 9.c].

Instead, when the quadrupole is in capacitive contact with the subjacent medium, and so the ratio $x=h/L$ between its height $h$ above ground and its electrode-electrode distance $L$ is not null, i.e. $0<x\leq1$, then the quadrupole is characterized by a geometrical factor $K(x)$ $[\delta(x)]$, decreasing (increasing) function of $x$, which, in the square configuration, slopes down (up) more swiftly than the Wenner arrangement, so assuming smaller (larger) values especially for $1/2<x<1$ [fig. 8.a]. As a consequence, a probe with a fixed $L$, which performs measurements on a medium of dielectric permittivity $\varepsilon_r$, could be designed with an optimal height/dimension ratio $x_{opt}=h_{opt}/L$ which, in the square configuration, is smaller than the Wenner arrangement, because its factor $\delta(x)$ slopes up more swiftly increasing the ratio $x$, so reaching the prefixed optimal value $\delta_{opt}(\varepsilon_r) \approx 2/(15\varepsilon_r+17)$ in correspondence with a smaller $x_{opt}$. In simpler terms, if the probe is in capacitive contact with the medium, in order to perform optimal measurement of the permittivity, then the square configuration needs to be raised above ground less than the Wenner arrangement, their electrode-electrode distance being equal. In fact, $x$ ranges from $x_{W,low}=0.022$ in the linear configuration and from $x_{S,low}=0.019$ in the square arrangement.

Moreover, in the case of capacitive contact, if the quadrupole, with electrode-electrode distance $L$, is designed to the optimal height/dimension ratio $x_{opt}=h_{opt}/L$, working in a frequency $f$, then the transfer impedance $Z(f,x_{opt})$ in units of $1/h_{opt}$, calculated in $x_{opt}$, is defined by a phase $\Phi(f,x_{opt})$, which does not depend on the square or Wenner configurations [fig. 8.d], and a modulus $|Z|(f,x_{opt})$ in units of $1/h_{opt}$, which, in the square is shifted down by a factor $1/\alpha$ with respect to the Wenner configuration [fig. 8.c], maintaining almost unvaried in both configurations not only the shape of the modulus $|Z|(f,x_{opt})$ but also the position of its zero $z(x_{opt})$ and pole $p(x_{opt})$ frequencies [fig. 8.b]

Finally, the inaccuracies $\Delta\sigma/\sigma(f,x_{opt})$ in the measurement of the conductivity $\sigma$ and $\Delta\varepsilon_r/\varepsilon_r(f,x_{opt})$ for the permittivity $\varepsilon_r$, calculated in $x_{opt}$, do not depend on the two configurations, so the optimal frequency $f_{opt}(x_{opt})$, which minimizes the inaccuracy $\Delta\varepsilon_r/\varepsilon_r(f,x_{opt})$ for $\varepsilon_r$, together with the minimum and maximum values of frequency, respectively $f_{min}(x_{opt})$ and $f_{max}(x_{opt})$, allowing the inaccuracy $\Delta\varepsilon_r/\varepsilon_r(f,x_{opt})$ below a prefixed limit (*10%*), are invariant in both the configurations [fig. 8.e]. In simpler words, if the probe is in capacitive contact with the medium, in order to perform an optimal measurement of permittivity, then the design of the two configurations establishes, as regards a different height/dimension ratio, (almost) invariant trends in frequency, both for their transfer impedances and measurement inaccuracies .

# 6. Conclusions.

The present paper proposed a theoretical modelling of the simultaneous and non invasive measurement of electrical resistivity and dielectric permittivity, using a quadrupole probe on a subjacent medium. A mathematical-physical model has been applied on propagation of errors in the measurement of resistivity and permittivity based on the sensitivity functions tool. The findings have also been compared to the results of the classical method of analysis in the frequency domain, which is useful for determining the behaviour of zero and pole frequencies in the linear time invariant (LTI) circuit of the quadrupole. This paper underlined that average values of electrical resistivity and dielectric permittivity may be used to estimate the complex impedance over various terrains and concretes, especially when they are characterized by low levels of water saturation or content (Knight and Nur, 1987) and analyzed within a bandwidth ranging only from low (LF) to middle (MF) frequencies (Myounghak et al., 2007)(Al-Qadi et al., 1995). In order to meet the design specifications which ensure satisfactory performances of the probe (inaccuracy no more than *10%*), the forecasts provided by the theory of error propagation [suggested by (Vannaroni et al., 2004) applying the sensitivity functions approach, explicitly developed in the paper, are less stringent than those foreseen by the analysis in the frequency domain [suggested by (Grard and Tabbagh, 1991)], deepening here the transfer function method to analyze the zero and pole behaviour (in terms of both a larger band of frequency $f$ and a wider measurable range of resistivity $\rho$ or permittivity $\varepsilon_r$).

It is interesting to compare the results of the present paper with those published in scientific literature (Grard and Tabbagh, 1991)(Vannaroni et al., 2004). In accordance, the sensitivity functions approach, provides the following results: a) if the quadrupole probe is in galvanic contact with the subsurface, i.e. *h=0*, then the inaccuracies $\Delta\sigma/\sigma$ in the measurement of conductivity $\sigma$ and $\Delta\varepsilon_r/\varepsilon_r$ for permittivity $\varepsilon_r$ are minimized in the frequency band $B$ of the quadrupole, for all its geometric configurations and media; b) and, even if *h≠0,* the design of the probe must be optimized with reference to the minimum value of the inaccuracy $\Delta\varepsilon_r/\varepsilon_r$ for $\varepsilon_r$, which is always higher than the corresponding minimum value of the inaccuracy $\Delta\sigma/\sigma$ in the band $B$, for all its configurations and media.

More explicitly than in referred papers, the transfer functions method provides results for which, in order to satisfy the operative conditions of linearity for the measurements: a) if the quadrupole has a capacitive contact with the subjacent medium, then one should impose the frequency $f$ of the probe to be included between the zero $z_M$ and the pole $p_M$ of the transfer impedance, so its modulus to be almost constant within the frequency band, so an optimization equation is deduced for the probe, which links the optimal ratio $x$ between its height above ground and its characteristic geometrical dimension only to the dielectric permittivity $\varepsilon_r$ of the medium; b) instead, if the quadrupole is in galvanic contact with the subjacent medium, then one should impose the working frequency $f$ of the quadrupole to be lower than the cut-off frequency of the transfer impedance, so its modulus to be constant below the cut-off frequency, so it is optimal to design the characteristic geometrical dimensions of the probe or establish the measurable ranges of the conductivity $\sigma$ and permittivity $\varepsilon_r$ of the medium.

Unlike referred papers, the sensitivity functions approach and the transfer functions method provide results which permit an assessment of the performance of the quadrupole probe in galvanic and capacitive contact: a) usually, having selected the surface (for example, a non-saturated concrete with low conductivity $\sigma=10^{-4}$ S/m and $\varepsilon_r=4$), it is possible to verify that the quadrupole in capacitive contact performs optimal measurements over the band $[f_{min}(x_{opt})<z(x_{opt}), f_{max}(x_{opt})>p(x_{opt})]$, which is shifted to lower and higher frequencies compared to the case when the probe is in galvanic contact, being the corresponding band $[f_{min}, f_{max}]$ narrower of almost one decade in frequency, especially increasing the value of $\varepsilon_r$; b) usually, having fixed the frequency band, the quadrupole in capacitive contact provides optimal measurements over surfaces of lower conductivity compared to when the probe is in galvanic contact, being the respective conductivities higher even of almost one magnitude order.

On this basis, some constraints were established to design a quadrupole probe for conducting measurements of electrical resistivity and dielectric permittivity in a regime of alternating current at low and middle frequencies (*10kHz-1MHz*). Measurement is carried out using four electrodes laid on the surface to be analyzed and, through a measurement of transfer impedance, there is the possibility of extracting the resistivity and permittivity of the material. Furthermore, increasing the distance between the electrodes, it is possible to investigate the electrical properties of the sub-surface structures to greater depth. The main advantage of the quadrupole is being able to conduct measurements of electrical parameters with a non destructive technique, thereby enabling characterization of precious and unique materials. Also, in appropriate arrangements, measurements could be carried out with electrodes slightly raised above the surface, enabling completely non-destructive analysis, although with a greater error. The probe is able to perform measurements on materials with high resistivity and permittivity in an immediate way, without subsequent stages of post-analysis of data.

## Appendix A.

There follows a discussion of the influence of the inaccuracies in transfer impedance in modulus and phase on the measurement of electrical conductivity and dielectric permittivity. The mathematical tool best suited to this purpose applies the so-called sensitivity functions (Murray-Smith, 1987), which formalize the intuitive concept of sensitivity as the ratio between the percentage error of certain physical quantities (due to the variation of some parameters) and the percentage error of the same parameters.

The inaccuracies $\Delta\sigma/\sigma$, in the measurement of the electrical conductivity $\sigma$, and $\Delta\varepsilon_r/\varepsilon_r$, for the dielectric permittivity $\varepsilon_r$, can be expressed as linear combinations of the inaccuracies $\Delta|Z|/|Z|$ and $\Delta\Phi_Z/\Phi_Z$ in the measurement of transfer impedance, respectively in modulus $|Z|$ and in phase $\Phi_Z$, as reported in eqs. (4.1) and (4.2) [figs. 3]. The pairs of sensitivity functions ($S_\sigma^{|Z|}, S_\sigma^{\Phi_Z}$) and ($S_{\varepsilon_r}^{|Z|}, S_{\varepsilon_r}^{\Phi_Z}$) for the transfer impedance, both in $|Z|$ and $\Phi_Z$, relative to the conductivity $\sigma$ and the permittivity $\varepsilon_r$ [figs. 4],

$$S_\sigma^{|Z|} = \frac{\Delta|Z|/|Z|}{\Delta\sigma/\sigma} = \frac{\partial|Z|}{\partial\sigma}\frac{\sigma}{|Z|} = $$
$$= \frac{1}{S_{|Z|}^\sigma} = \frac{1}{2}H_1 S_\sigma^R - \frac{1}{2}H_2 S_\sigma^C \quad, \quad \text{for} \quad \varepsilon_r = const \tag{A.1}$$

$$S_\sigma^{\Phi_Z} = \frac{\partial \Phi_Z}{\partial \sigma}\frac{\sigma}{\Phi_Z} =$$
$$= \frac{1}{S_{\Phi_Z}^\sigma} = \frac{1}{2}\frac{\Omega\frac{R}{R_N}\frac{C}{C_N}}{arct(\Omega\frac{R}{R_N}\frac{C}{C_N})}H_1(S_\sigma^R + S_\sigma^C) \cong \frac{1}{2}H_1(S_\sigma^R + S_\sigma^C) \;, for \; \Omega\frac{R}{R_N}\frac{C}{C_N} \leq 1 \; (\varepsilon_r = const) \tag{A.2}$$

$$S_{\varepsilon_r}^{|Z|} = \frac{1}{S_{|Z|}^{\varepsilon_r}} = \frac{1}{2}H_1 S_{\varepsilon_r}^R - \frac{1}{2}H_2 S_{\varepsilon_r}^C \quad, \quad \text{for} \quad \sigma = const, \tag{A.3}$$

$$S_{\varepsilon_r}^{\Phi_Z} = \frac{1}{S_{\Phi_Z}^{\varepsilon_r}} = \frac{1}{2}\frac{\Omega\frac{R}{R_N}\frac{C}{C_N}}{arct(\Omega\frac{R}{R_N}\frac{C}{C_N})}H_1(S_{\varepsilon_r}^R + S_{\varepsilon_r}^C) \cong \frac{1}{2}H_1(S_{\varepsilon_r}^R + S_{\varepsilon_r}^C) \;, for \; \Omega\frac{R}{R_N}\frac{C}{C_N} \leq 1 \; (\sigma = const), \tag{A.4}$$

are, in turn, linear combinations of the sensitivity function pairs ($S_\sigma^R, S_\sigma^C$) and ($S_{\varepsilon_r}^R, S_{\varepsilon_r}^C$) for transfer impedance, in both the resistance $R$ and capacitance $C$ parallel components, relative to $\sigma$ and $\varepsilon_r$,

$$S_\sigma^R = \left.\frac{\Delta R/R}{\Delta\sigma/\sigma}\right|_{\varepsilon_r=const} = \left.\frac{\partial R}{\partial \sigma}\right|_{\varepsilon_r=const}\frac{\sigma}{R} = -\frac{[1+\delta(x)\frac{\varepsilon_r-1}{2}]^2 - [\frac{\delta(x)}{\Omega}\frac{\varepsilon_r+1}{2}]^2}{[1+\delta(x)\frac{\varepsilon_r-1}{2}]^2 + [\frac{\delta(x)}{\Omega}\frac{\varepsilon_r+1}{2}]^2}, \tag{A.5}$$

$$S_\sigma^C = \left.\frac{\partial C}{\partial \sigma}\right|_{\varepsilon_r=const}\frac{\sigma}{C} = \frac{1}{2}\frac{\varepsilon_r+1}{\Omega^2}\frac{\delta(x)[2+\delta(x)(\varepsilon_r-3) - \delta^2(x)(\varepsilon_r-1)]}{[1+\delta(x)(\frac{\varepsilon_r-1}{2}+\frac{1}{\Omega^2}\frac{\varepsilon_r+1}{2})][(1+\delta(x)\frac{\varepsilon_r-1}{2})^2 + (\frac{\delta(x)}{\Omega}\frac{\varepsilon_r+1}{2})^2]}, \tag{A.6}$$

$$\left.S_{\varepsilon_r}^R\right|_{\sigma=const} = \varepsilon_r \frac{\delta(x)[1+\delta(x)\frac{\varepsilon_r-1}{2}]}{[1+\delta(x)\frac{\varepsilon_r-1}{2}]^2 + [\frac{\delta(x)}{\Omega}\frac{\varepsilon_r+1}{2}]^2}, \tag{A.7}$$

$$S^C_{\varepsilon_r}\Big|_{\sigma=const} = \frac{\varepsilon_r}{\varepsilon_r+1} \frac{1+\delta(x)(\varepsilon_r-2)+\frac{1}{4}\delta^2(x)[(\varepsilon_r-1)(\varepsilon_r-5)-(\frac{\varepsilon_r+1}{\Omega})^2]-\frac{1}{4}\delta^3(x)[(\varepsilon_r-1)^2-(\frac{\varepsilon_r+1}{\Omega})^2]}{[1+\delta(x)(\frac{\varepsilon_r-1}{2}+\frac{1}{\Omega^2}\frac{\varepsilon_r+1}{2})][(1+\delta(x)\frac{\varepsilon_r-1}{2})^2+(\frac{\delta(x)}{\Omega}\frac{\varepsilon_r+1}{2})^2]}, \quad (A.8)$$

with the weight functions

$$H_1 = \frac{[1-\delta(x)]^2}{1+\Omega^2+\delta(x)(\varepsilon_r-1)(1+\Omega^2)+\delta^2(x)(1+\Omega\frac{\varepsilon_r-1}{2}+\frac{1}{\Omega}\frac{\varepsilon_r+1}{2})^2}, \quad (A.9)$$

$$H_2 = \frac{\Omega^2+2\delta(x)(\Omega^2\frac{\varepsilon_r-1}{2}+\frac{\varepsilon_r+1}{2})+\delta^2(x)(\Omega\frac{\varepsilon_r-1}{2}+\frac{1}{\Omega}\frac{\varepsilon_r+1}{2})^2}{1+\Omega^2+\delta(x)(\varepsilon_r-1)(1+\Omega^2)+\delta^2(x)(1+\Omega\frac{\varepsilon_r-1}{2}+\frac{1}{\Omega}\frac{\varepsilon_r+1}{2})^2}. \quad (A.10)$$

Discussing eqs. (A.1)-(A.4), if the modulus $|Z|$ and the phase $\Phi_Z$ of the transfer impedance provide an indirect measurement of the electrical conductivity $\sigma$ and dielectric permittivity $\varepsilon_r$, then the functions $|Z|=|Z|(\sigma, \varepsilon_r)$ and $\Phi_Z=\Phi_Z(\sigma, \varepsilon_r)$ are invertible, i.e. $\sigma=\sigma(|Z|, \Phi_Z)$ and $\varepsilon_r=\varepsilon_r(|Z|, \Phi_Z)$. Therefore, the theorem of the derivative for the inverse function can be applied. In fact, under the condition $\sigma=const$ (or $\varepsilon_r=const$), both $|Z|$ and $\Phi_Z$ are invertible functions of $\varepsilon_r$ (or $\sigma$), i.e. strictly increasing or decreasing monotonic functions of $\varepsilon_r$ (or $\sigma$).

## Appendix B.

By exact calculation, the transfer impedance $Z(f,x,\sigma,\varepsilon_r)$ measured by the quadrupole probe, in units of the reciprocal height $1/h$ from the subjacent medium, consists of the resistance $R(f,x,\sigma,\varepsilon_r)$, in units of $1/h$ [see eq. (3.10)], which can be expressed as a transfer function characterized by a pole in the origin frequency, $p_R=0$, a zero in higher frequencies $z_R(f,x,\sigma,\varepsilon_r)>0$, and a static gain $K_R(f,x,\sigma)$,

$$R(f,x,\sigma,\varepsilon_r) = K_R(x,\sigma)\frac{1+\frac{f^2}{z_R^2(x,\sigma,\varepsilon_r)}}{(2\pi f)^2}, \quad (B.1)$$

where

$$z_R(x,\sigma,\varepsilon_r) = \frac{1}{2\pi} \frac{\sigma}{\varepsilon_0(\varepsilon_r+1)} \frac{\delta(x)\frac{\varepsilon_r+1}{2}}{1+\delta(x)\frac{\varepsilon_r-1}{2}}, \tag{B.2}$$

$$K_R(x,\sigma) = \frac{1}{2}\frac{\sigma}{\varepsilon_0}\frac{1}{C_0(x)}\frac{\delta^2(x)}{1-\delta(x)}, \tag{B.3}$$

besides the parallel capacitance $C(f,x,\sigma,\varepsilon_r)$, in units of $1/h$ [see eq. (3.11)], which can be expressed as a transfer function characterized by a low frequency pole, $p_C(f,x,\sigma,\varepsilon_r)$, a zero in higher frequencies $z_C(f,x,\sigma,\varepsilon_r) > p_C(f,x,\sigma,\varepsilon_r)$, and a static gain $K_C(x)$,

$$C(f,x,\sigma,\varepsilon_r) = K_C(x) \frac{1+\frac{f^2}{z_C^2(x,\sigma,\varepsilon_r)}}{1+\frac{f^2}{p_C^2(x,\sigma,\varepsilon_r)}}, \tag{B.4}$$

where the capacitance pole $p_C(f,x,\sigma,\varepsilon_r)$ coincides with the resistance pole $z_R(f,x,\sigma,\varepsilon_r)$,

$$p_C(x,\sigma,\varepsilon_r) = z_R(x,\sigma,\varepsilon_r), \tag{B.5}$$

and

$$z_C(x,\sigma,\varepsilon_r) = \frac{1}{2\pi}\frac{\sigma}{\varepsilon_0(\varepsilon_r+1)}\sqrt{\frac{\delta(x)\frac{\varepsilon_r+1}{2}}{1+\delta(x)\frac{\varepsilon_r-1}{2}}}, \tag{B.6}$$

$$K_C(x) = \frac{C_0(x)}{\delta(x)}. \tag{B.7}$$

One can demonstrate that, for values of the ratio $x=h/L$, between the height $h$ above ground and the characteristic geometrical dimension $L$, and the paired values of electrical conductivity $\sigma$ and dielectric permittivity $\varepsilon_r$ which satisfy the condition [fig. 5]

$$\frac{1}{[K_R(x,\sigma)K_C(x)]^2} << \frac{2}{[2\pi \cdot z_C(x,\sigma,\varepsilon_r)]^2}, \tag{B.8}$$

the modulus $|Z|(f,x,\sigma,\varepsilon_r)$ can be approximately expressed as a transfer function with a pole in the origin frequency, a low frequency zero, $z_M(f,x,\sigma,\varepsilon_r)$, a pole in higher frequencies $p_M(f,x,\sigma,\varepsilon_r) > z_M(f,x,\sigma,\varepsilon_r)$, and a static gain $K_M(x)$ [fig. 8.c],

$$|Z|(f,x,\sigma,\varepsilon_r) \cong K_M(x) \frac{1 + \frac{f^2}{z_M^2(x,\sigma,\varepsilon_r)}}{2\pi f \cdot [1 + \frac{f^2}{p_M^2(x,\sigma,\varepsilon_r)}]}, \tag{B.9}$$

where the zero of the modulus $z_M(f,x,\sigma,\varepsilon_r)$ coincides with the capacitance pole $p_C(f,x,\sigma,\varepsilon_r)$ and the pole of the modulus $p_M(f,x,\sigma,\varepsilon_r)$ with the capacitance zero $z_C(f,x,\sigma,\varepsilon_r)$ [fig. 8.b],

$$z_M(x,\sigma,\varepsilon_r) = p_C(x,\sigma,\varepsilon_r), \tag{B.10}$$

$$p_M(x,\sigma,\varepsilon_r) = z_C(x,\sigma,\varepsilon_r), \tag{B.11}$$

and

$$K_M(x) = \frac{1}{K_C(x)}. \tag{B.12}$$

Eq. (B.8) establishes limits on the range for the design specification $x$ of the quadrupole and the measurable range $(\sigma,\varepsilon_r)$ of the media.

In order to satisfy the operative conditions of linearity for the measurements, the quadrupole probe, characterized by the height/dimensions ratio $x=h/L$, should measure the conductivity $\sigma$ and the permittivity $\varepsilon_r$ of the subjacent medium when its working frequency $f$ falls within the band included between the zero $z_M(f,x,\sigma,\varepsilon_r)$ and the pole $p_M(f,x,\sigma,\varepsilon_r)$ of the transfer impedance, as reported in eq. (4.3).

Moreover, the quadrupole probe, specified by $x=h/L$, should measure $\varepsilon_r$, its geometric factor $\delta(x)$ being close to eq. (4.4), a necessary condition for $Z(f,x,\sigma,\varepsilon_r)$ to show an almost constant modulus within the band (4.3), the modulus in the zero (B.10) coinciding with the corresponding one in the pole (B.11),

$$|Z|_{f=z_M}(x,\sigma,\varepsilon_r) = \frac{K_M(x)}{\pi z_M(x,\sigma,\varepsilon_r)} \simeq |Z|_{f=p_M}(x,\sigma,\varepsilon_r) = \frac{1}{2} K_M(x) \frac{p_M(x,\sigma,\varepsilon_r)}{2\pi z_M^2(x,\sigma,\varepsilon_r)}, \tag{B.13}$$

so that the pole (B.11) is almost four times larger that the zero (B.10),

$$p_M(x,\sigma,\varepsilon_r) \approx 4 z_M(x,\sigma,\varepsilon_r). \tag{B.14}$$

Eq. (4.4) can be interpreted as the optimization equation of the quadrupole, so the sizing for the height/dimension ratio $x$ of the probe depends only on the permittivity $\varepsilon_r$ of the medium; instead, eqs. (4.3) and (B.14) show that the probe can work optimally only in a small band of frequencies.

**References.**

**Figures and captions.**

Figure 1

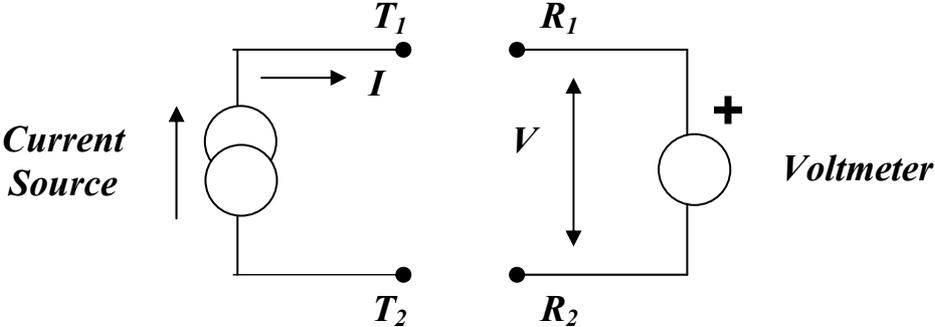
30

Figure 2.a

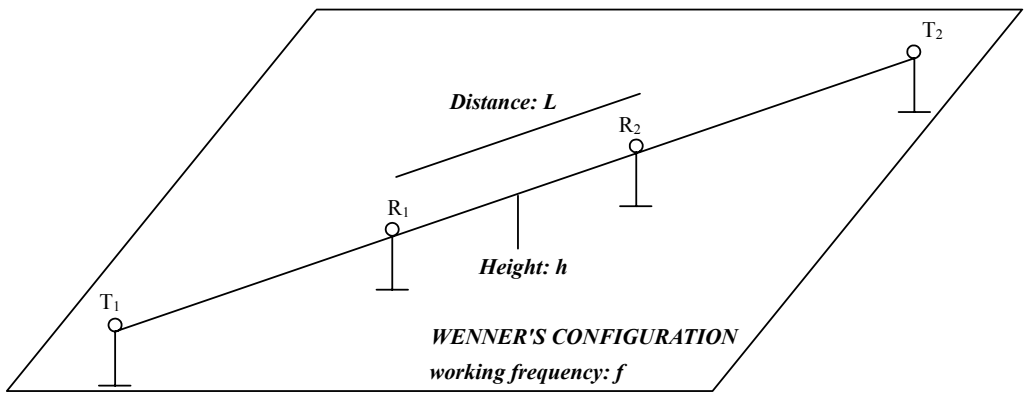

Figure 2.b

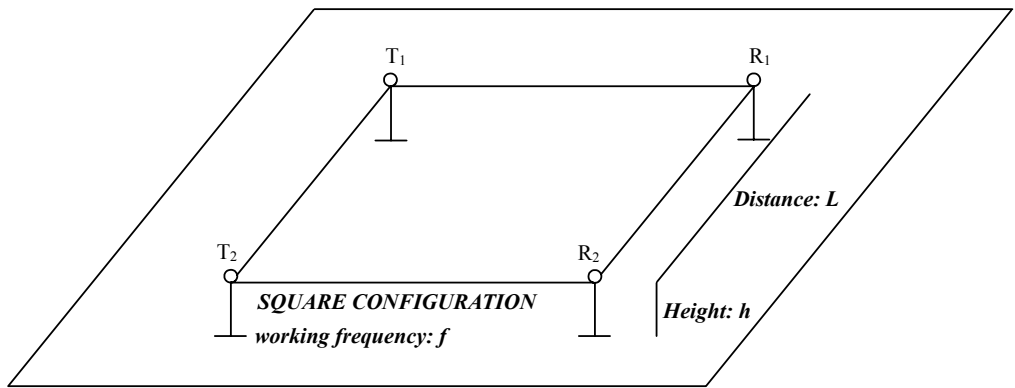

Figure 3.a

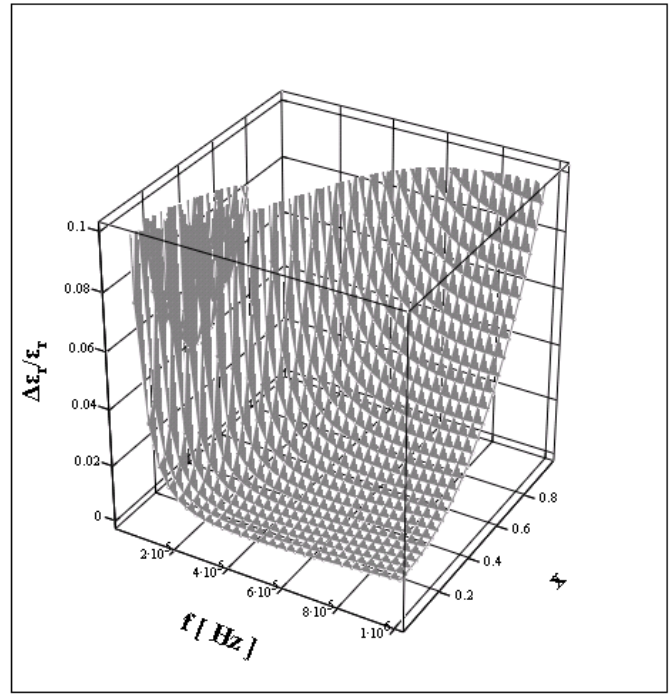

Figure 3.a.bis

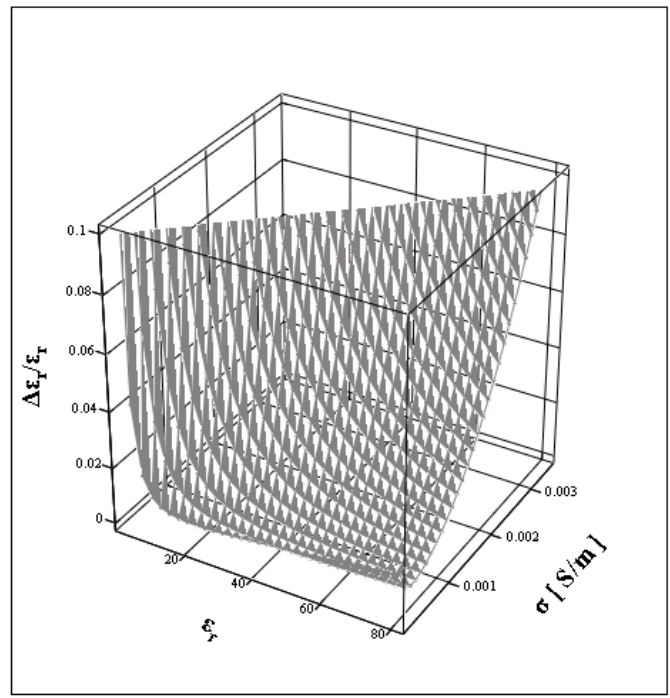

Figure 4.a

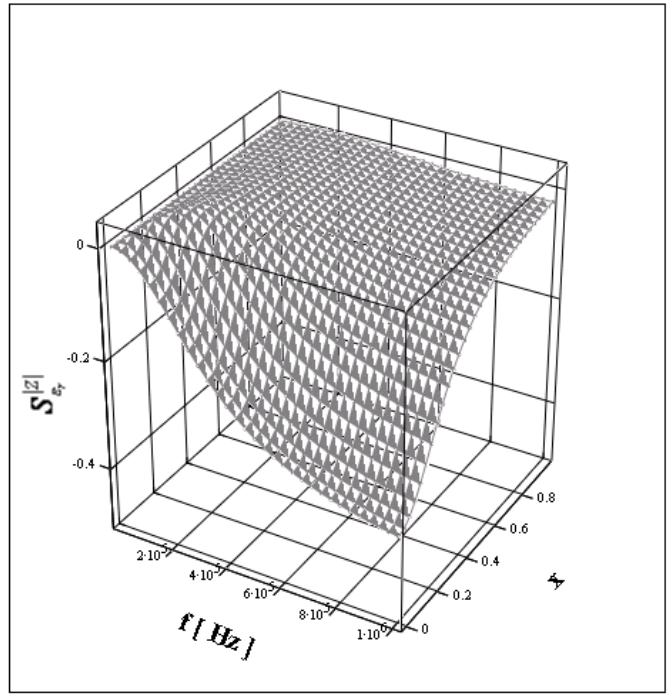

Figure 4.a.bis

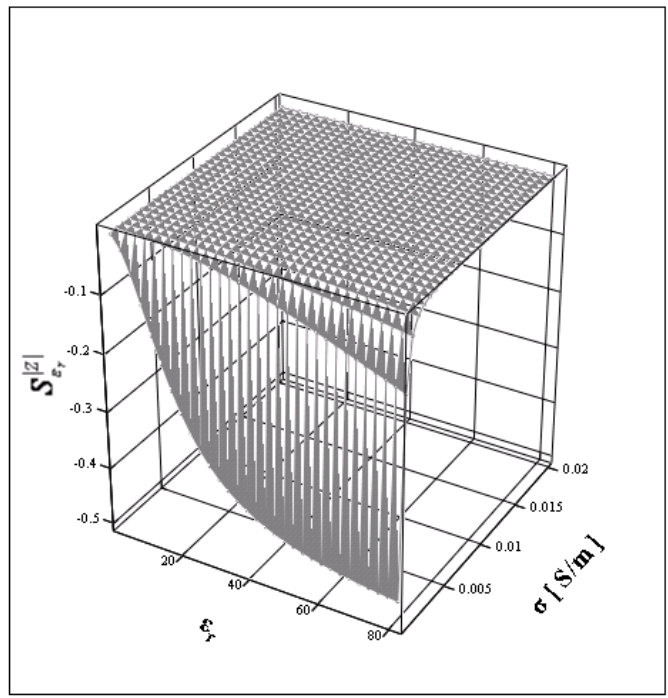

Figure 4.b

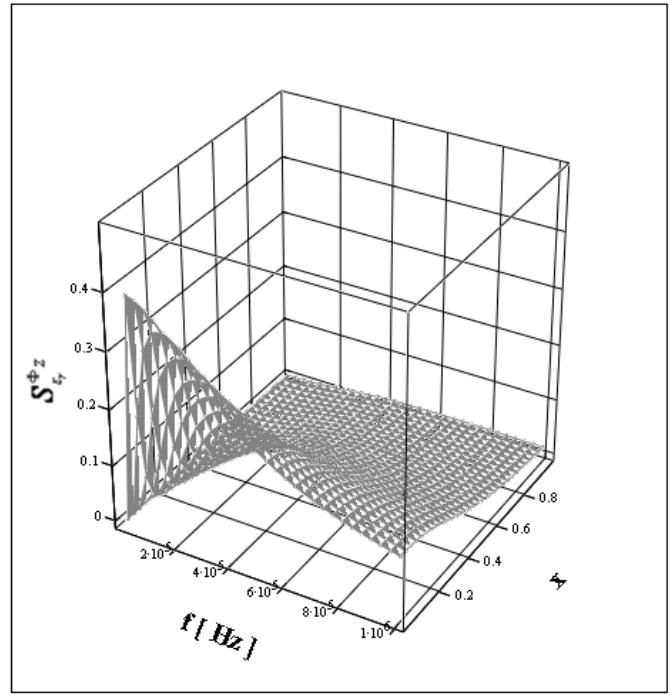

Figure 4.b.bis

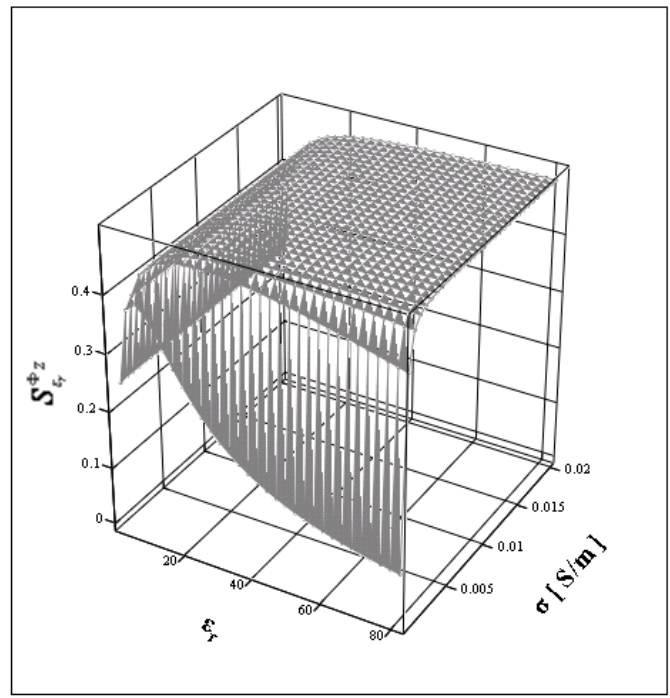

Figure 5

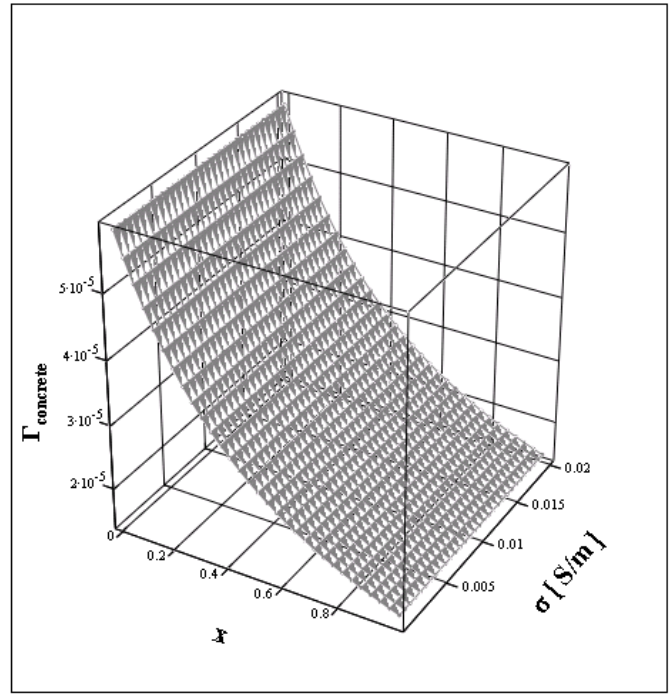

Figure 6.a.

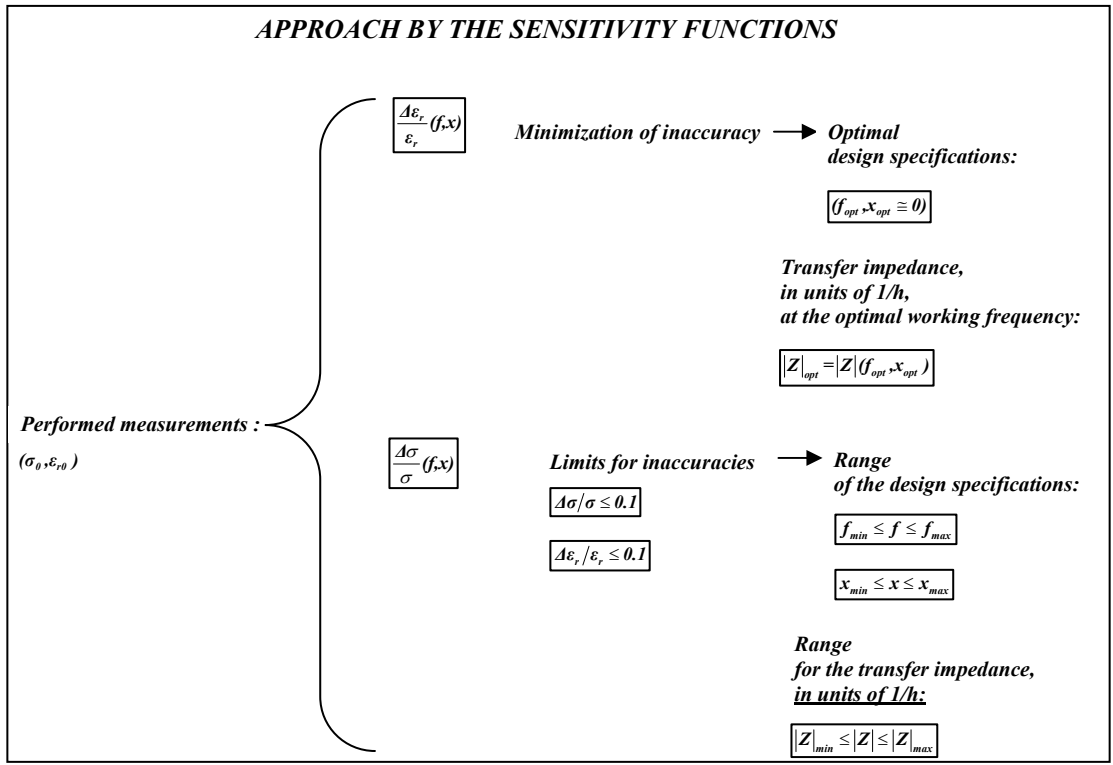

Figure 6.b.

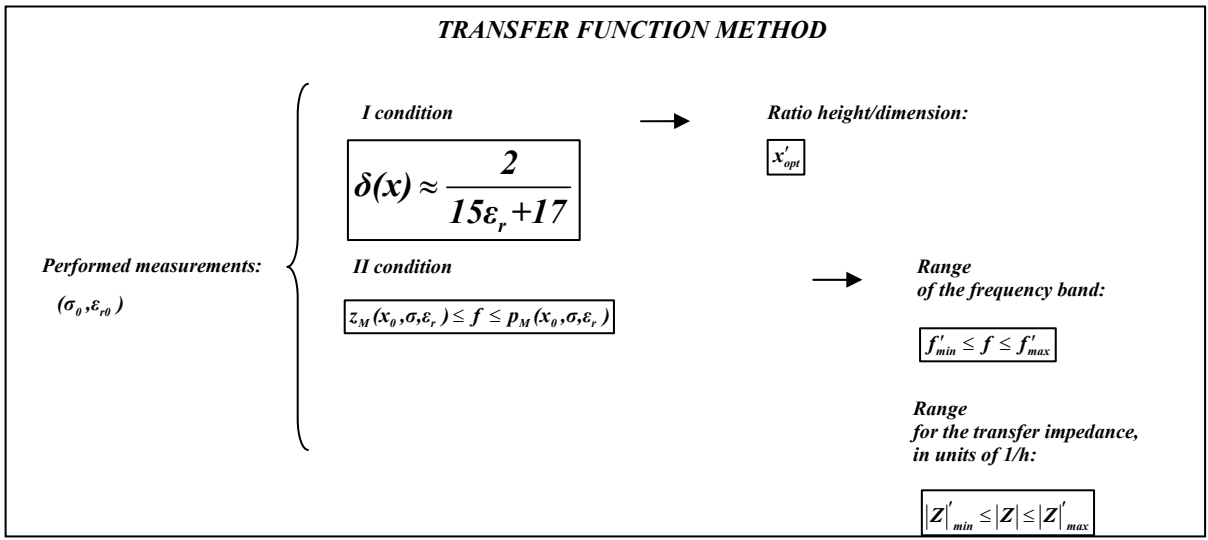

Figure 7.a.

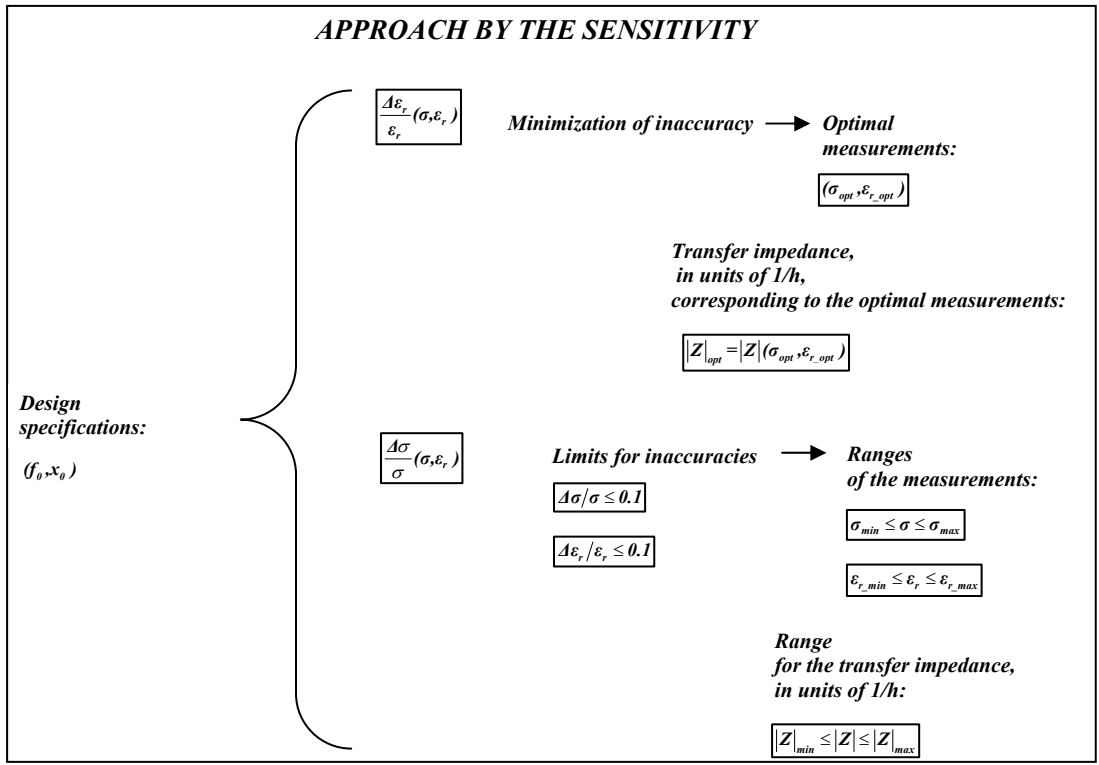

Figure 7.b.

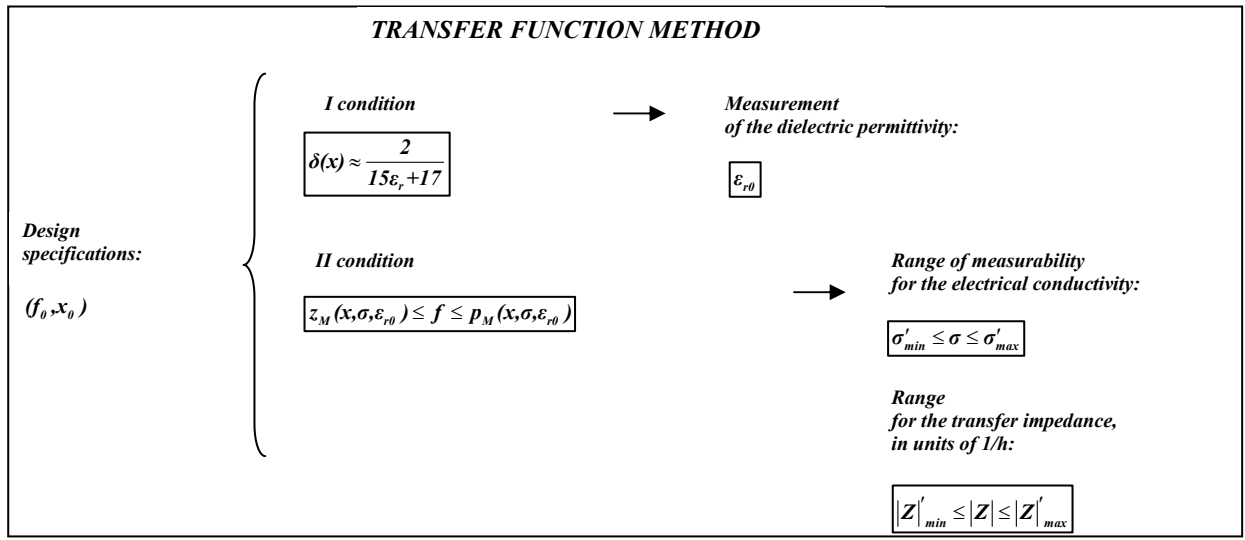

Figure 8.a

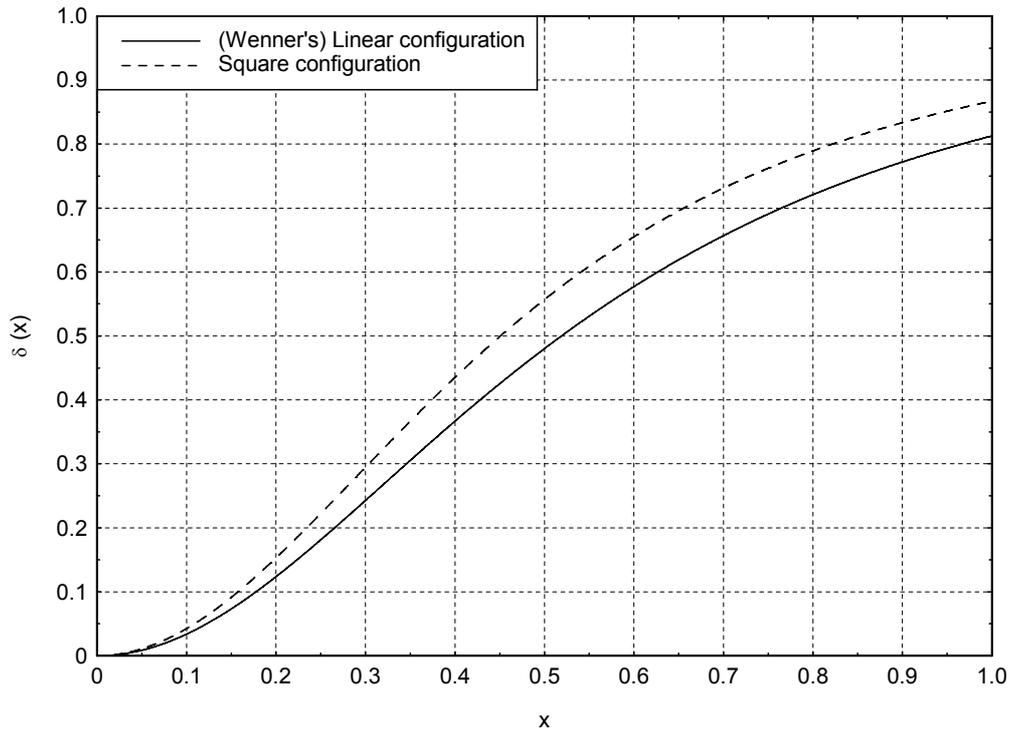

Figure 8.b

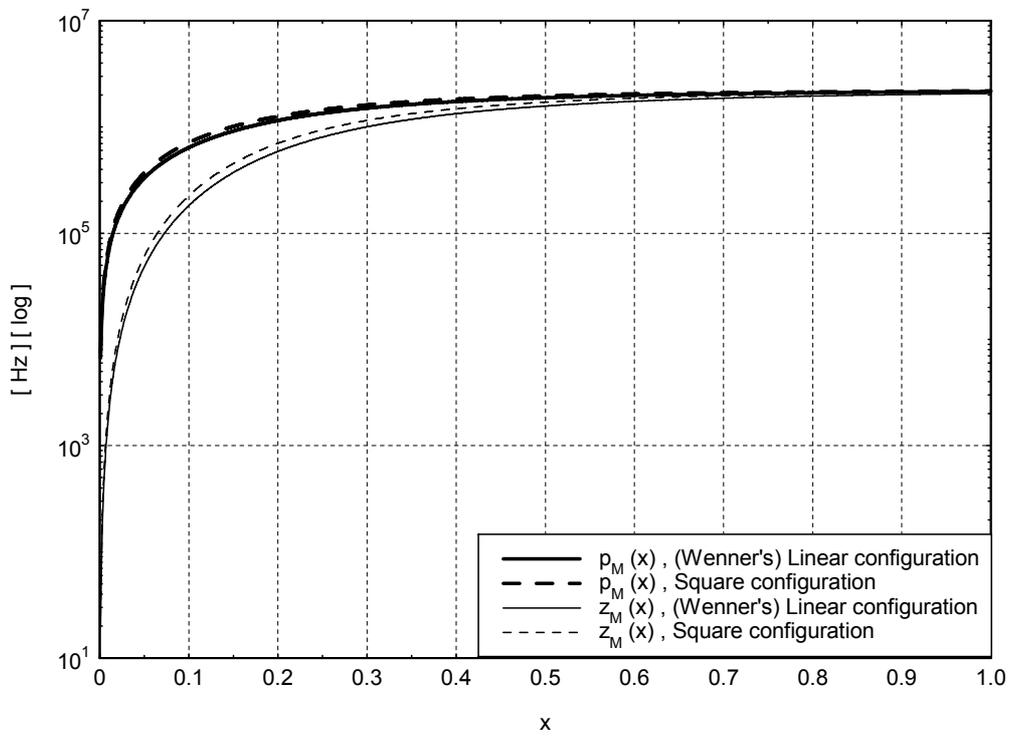

Figure 8.c

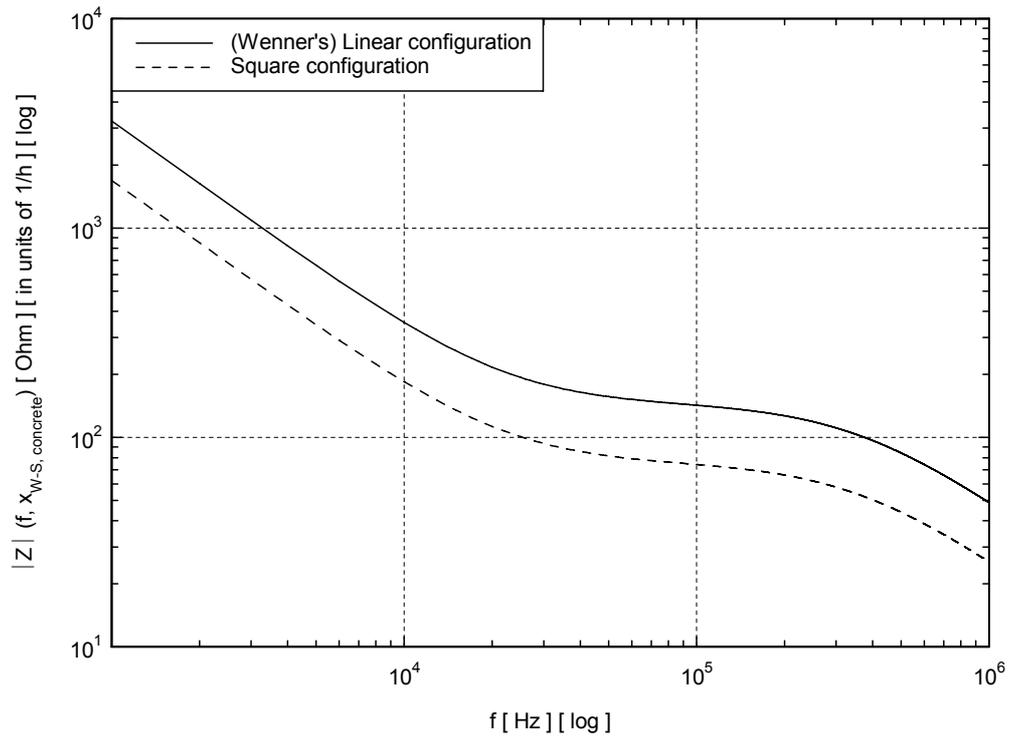

Figure 8.d

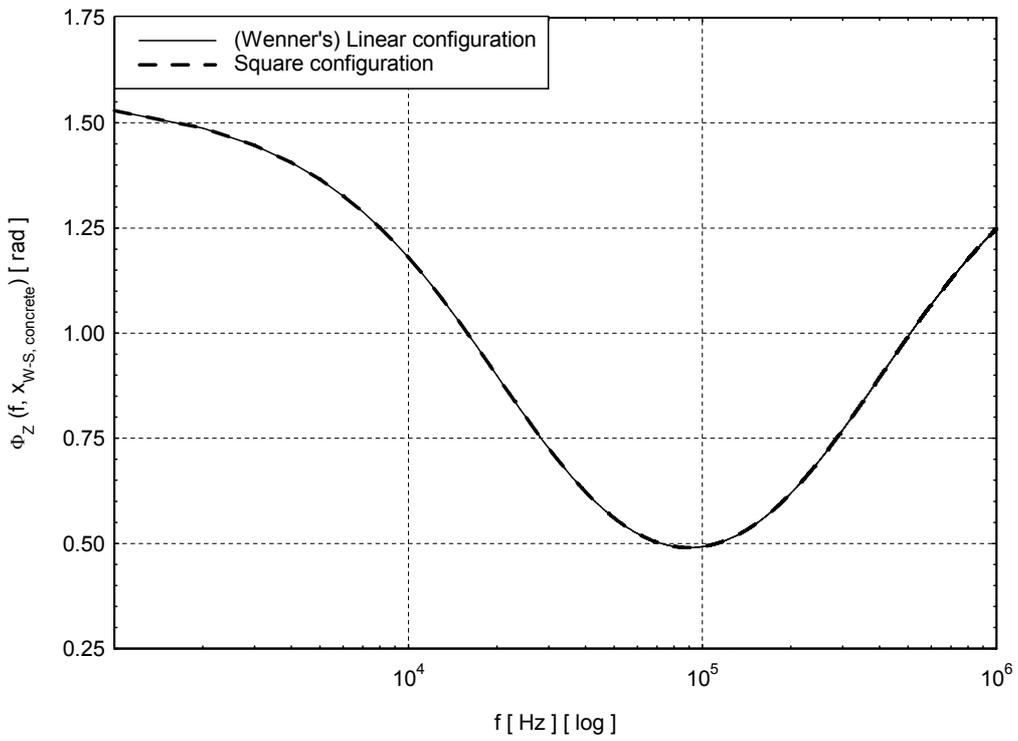

Figure 8.e

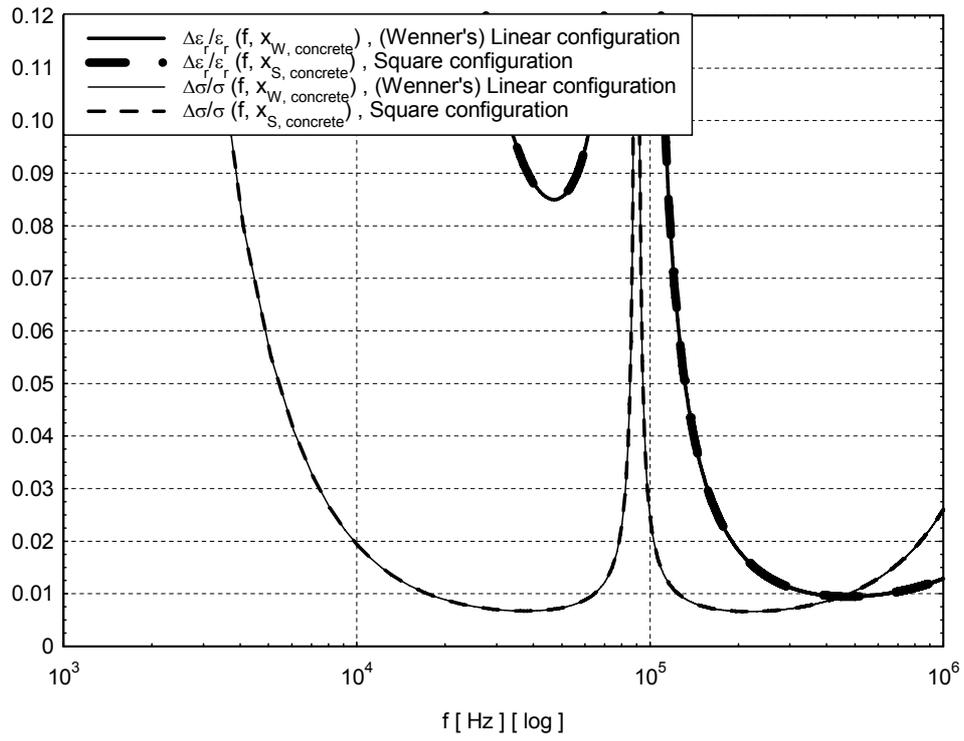

Figure 9.a

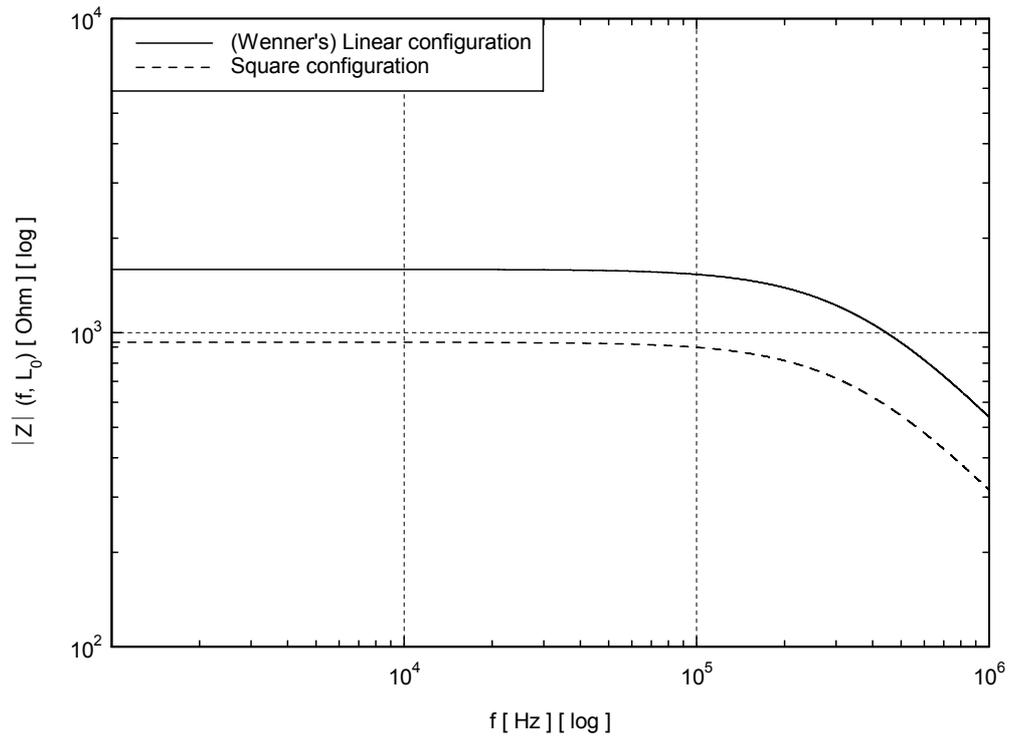

Figure 9.b

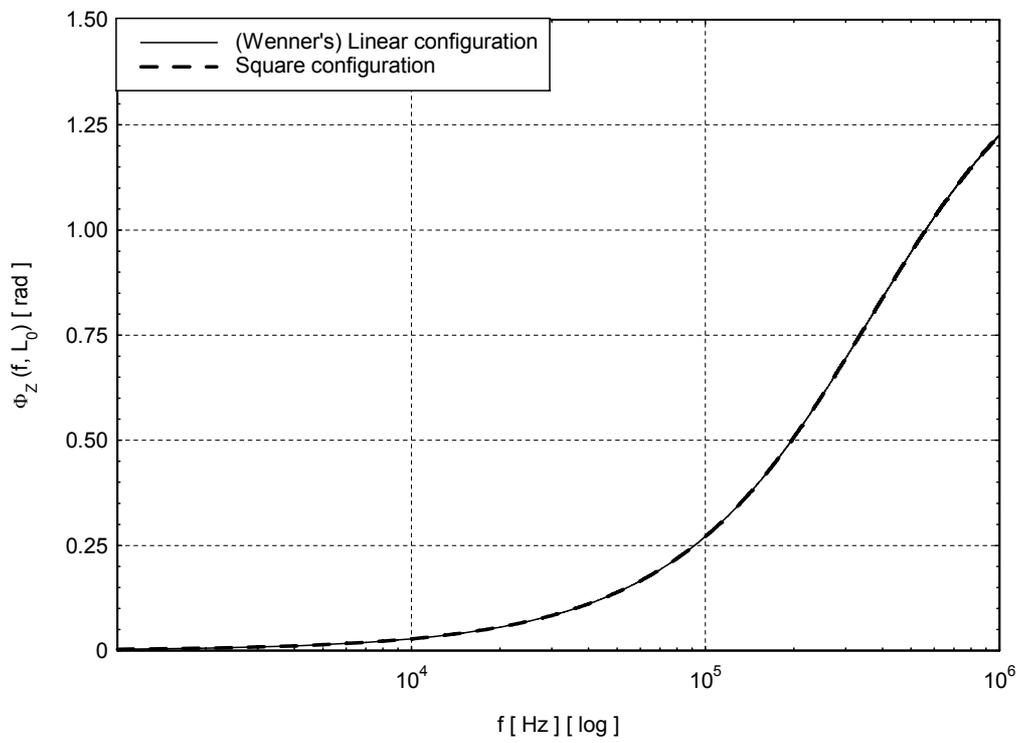

Figure 9.c

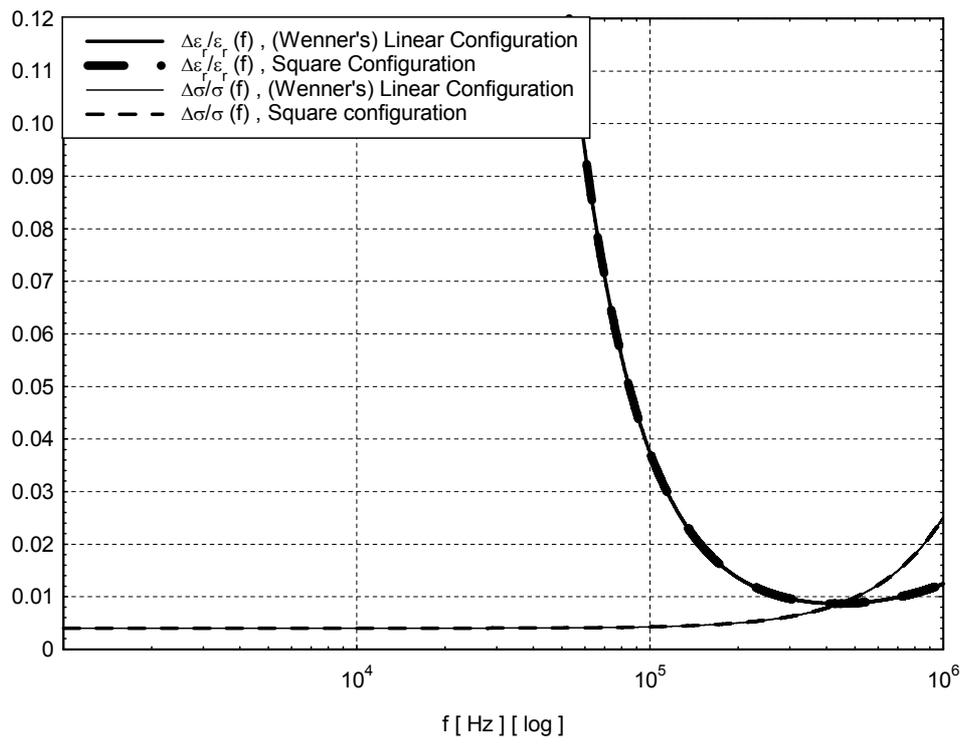

Figure 10.a

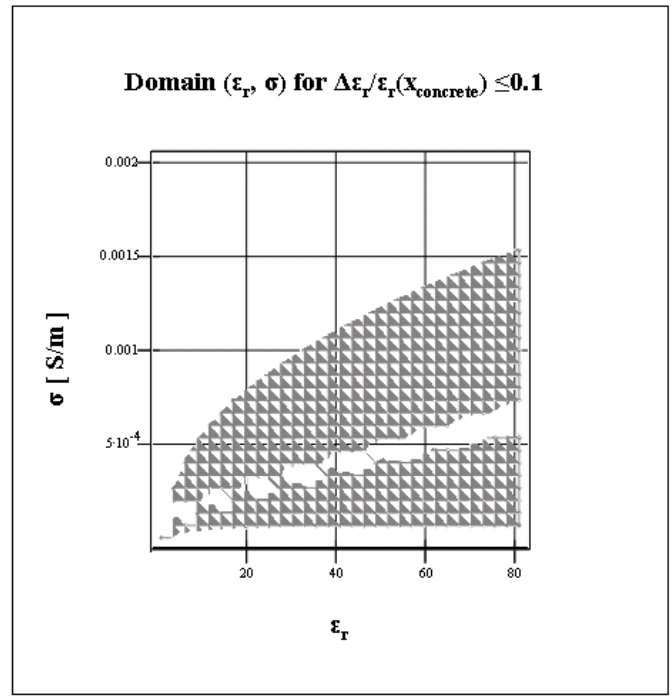

Figure 10.b

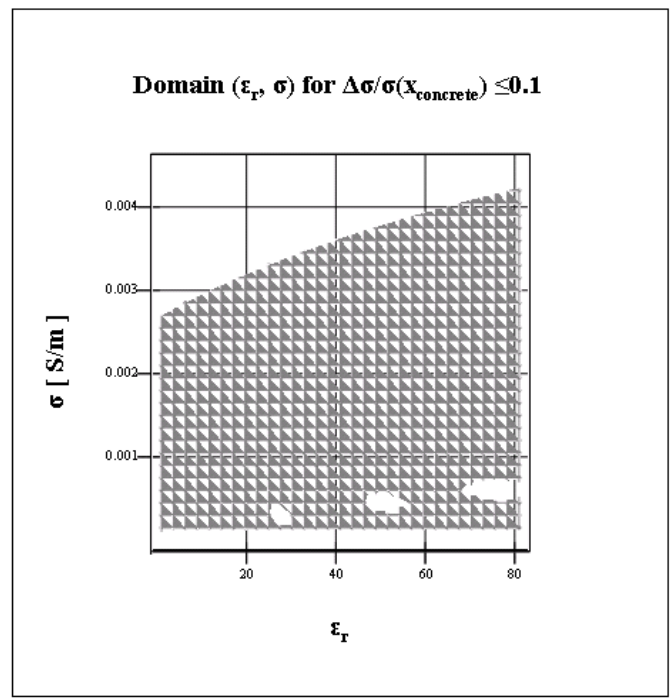

Figure 11.a

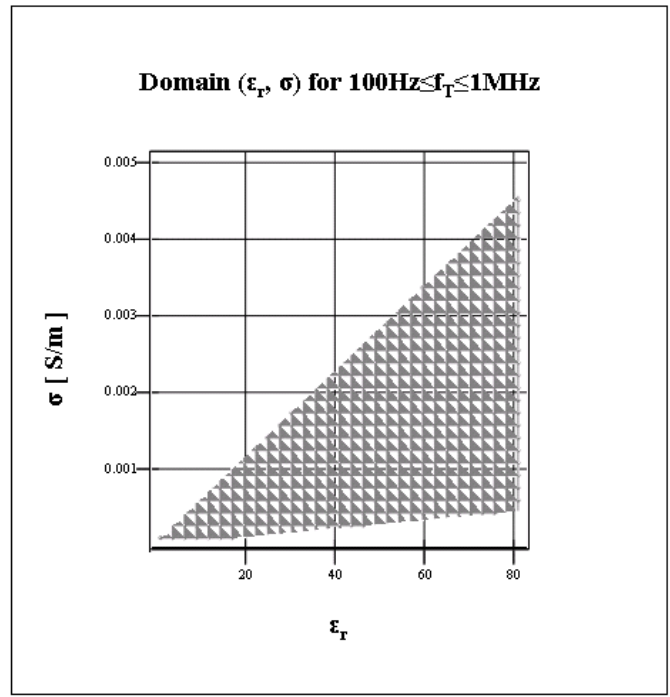

Figure 11.b

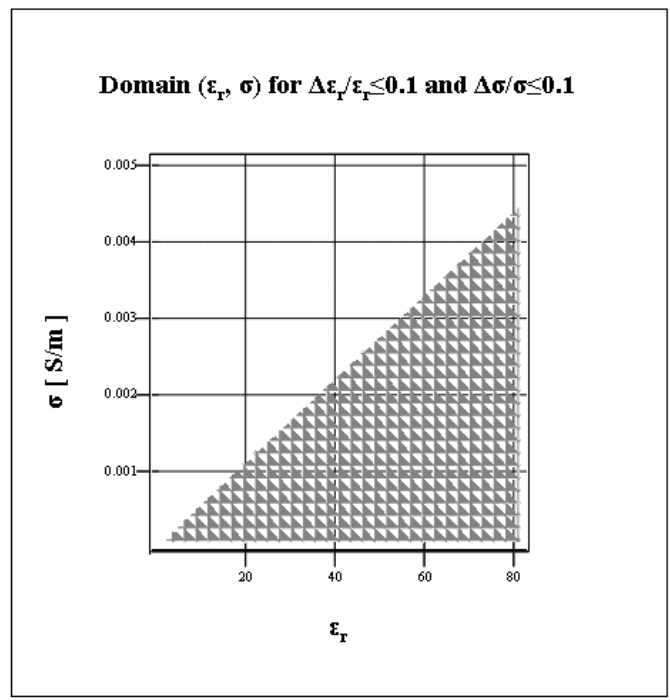

Table 1.a

|  | **Permittivity Inaccuracy** |
|---|---|
| $x_{W,\ opt}$ | $1.083 \cdot 10^{-4}$ |
| $\varepsilon_{r,\ opt}$ | 6.703 |
| $\sigma_{opt}$ | $3.52 \cdot 10^{-5}$ S/m |

Table 1.b

| $\varepsilon_{r,\ opt}=6.703$ $\sigma_{opt}=3.52\cdot 10^{-5}$ S/m | $\Delta\varepsilon_r/\varepsilon_r \leq 0.1$ $\Delta\sigma/\sigma \leq 0.1$ |
|---|---|
| $x_{W,\ low}$ | $\approx 0$ |
| $x_{W,\ up}$ | 0.475 |

Table 1.c

| $x_{W,\ opt}=1.083\cdot 10^{-4}$ | $\Delta\varepsilon_r/\varepsilon_r \leq 0.1$ $\Delta\sigma/\sigma \leq 0.1$ |
|---|---|
| $\varepsilon_{r,\ low}$, $\sigma_{low}$ | 1, $5.333\cdot 10^{-5}$ S/m |
| $\varepsilon_{r,\ up}$, $\sigma_{up}$ | 81, $3.14\cdot 10^{-3}$ S/m |

Table 2.a

| $\varepsilon_{r,\,concrete}=4.026$<br>$x_{W,\,concrete}=0.087$ | **Sensitivity Function Approach** | **Transfer Function Method** |
|---|---|---|
| $\sigma_{low}$ | 4.473·10⁻⁶ S/m | 1.78·10⁻⁵ S/m |
| $\sigma_{up}$ | 3.058·10⁻⁴ S/m | 7.12·10⁻⁵ S/m |

Table 2.b

| $x_{W,\,concrete}=0.087$ | $\Delta\varepsilon_r/\varepsilon_r\leq 0.1$<br>$\Delta\sigma/\sigma\leq 0.1$ |
|---|---|
| $\varepsilon_{r,\,low}$, $\sigma_{low}$ | 1, 1.769·10⁻⁶ S/m |
| $\varepsilon_{r,\,up}$, $\sigma_{up}$ | 84.458, 1.573·10⁻³ S/m |

Table 3

| $h=0$ | $\Delta\varepsilon_r/\varepsilon_r\leq 0.1$<br>$\Delta\sigma/\sigma\leq 0.1$ |
|---|---|
| $\varepsilon_{r,\,low}$, $\sigma_{low}$ | 1, 5.333·10⁻⁵ S/m |
| $\varepsilon_{r,\,up}$, $\sigma_{up}$ | 81, 3.14·10⁻³ S/m |

Fig. 1. Equivalent circuit of the quadrupole probe.

Fig. 2. Quadrupole probe in linear (Wenner) (a) or square (b) configuration.

Fig. 3. In the hypothesis that $\Delta|Z|/|Z|=\Delta\Phi_Z/\Phi_Z=10^{-3}$, inaccuracy $\Delta\varepsilon_r/\varepsilon_r$ in the measurement of the dielectric permittivity $\varepsilon_r$, plotted as: function $\Delta\varepsilon_r/\varepsilon_r(f,x)$ (a) of both the frequency $f$ in the band $f \in [0, f_{lim}]$, being $f_{lim}=1MHz$, and the ratio $x=h/L$ between the height $h$ above ground and the characteristic geometrical dimension $L$, being $0<x\leq1$, when the quadrupole probe, designed in the Wenner linear configuration, has a capacitive contact on a selected concrete of low electrical conductivity, i.e. $\sigma=10^{-4}$ S/m, $\varepsilon_r=4$; function $\Delta\varepsilon_r/\varepsilon_r(\sigma,\varepsilon_r)$ (a.bis) of both the conductivity $\sigma$ and the permittivity $\varepsilon_r$, when the quadrupole, working in a fixed band $B=100kHz$, is in galvanic contact on a class of concretes such that $\sigma \in [10^{-4} S/m, 2\cdot10^{-2} S/m]$.

Fig. 4. Sensitivity functions $S_{\varepsilon_r}^{|Z|}$ and $S_{\varepsilon_r}^{\Phi_Z}$ for the transfer impedance, both in modulus $|Z|$ and in phase $\Phi_Z$, relative to the dielectric permittivity $\varepsilon_r$, plotted as: functions, $S_{\varepsilon_r}^{|Z|}(f,x)$ (a) and $S_{\varepsilon_r}^{\Phi_Z}(f,x)$ (b), of both the working frequency $f$ and the height/dimension ratio $x=h/L$ in the same operative conditions of fig.3.a; functions, $S_{\varepsilon_r}^{|Z|}(\sigma,\varepsilon_r)$ (a.bis) and $S_{\varepsilon_r}^{\Phi_Z}(\sigma,\varepsilon_r)$ (b.bis), of both the conductivity $\sigma$ and the permittivity $\varepsilon_r$ in the same operative conditions of fig. 3.a.bis.

Fig. 5. Ratio $\Gamma=\Gamma_1/\Gamma_2$ between the first member $\Gamma_1$ and the second member $\Gamma_2$ of eq. (B.8), plotted as function $\Gamma(x,\sigma)$ of both the height/dimension ratio $x=h/L$ and the electrical conductivity $\sigma$, being the quadrupole probe designed in the Wenner linear configuration and in capacitive contact on a selected concrete of dielectric permittivity $\varepsilon_r=4$.

Fig. 6. Conceptual schemes for the numerical simulations regarding the sensitivity functions approach (a) and the transfer function method (b), in order to design the characteristic geometrical dimensions and the frequency band, limiting inaccuracies in the measurements of the quadrupole probe, in capacitive contact with selected materials as concretes, in the hypothesis that $\Delta|Z|/|Z|=\Delta\Phi_Z/\Phi_Z=10^{-3}$.

Fig. 7. Conceptual schemes for the numerical simulations regarding the sensitivity functions approach (a) and the transfer function method (b), in order to establish the measurable ranges of electrical conductivity and dielectric permittivity, limiting inaccuracies in the measurements of the quadrupole probe, in capacitive contact, and fixing its optimum working frequencies and characteristic geometrical dimensions $[\Delta|Z|/|Z|=\Delta\Phi_Z/\Phi_Z=10^{-3}]$.

Fig. 8. With reference to a quadrupole probe designed in the Wenner linear or square configuration and presenting a capacitive contact on a concrete of low electrical conductivity, i.e. $\sigma=10^{-4}$ S/m, $\varepsilon_r=4$; plots, as function of the ratio $x=h/L$ between the height $h$ above ground and the characteristic geometrical dimension $L$, being $0<x\leq1$, for the geometrical factor $\delta(x)$ (a); semi-logarithmic plots for both the zero $z_M(x)$ and pole $p_M(x)$ of the transfer impedance in modulus (b); Bode's diagrams, as function of the frequency $f$ in the band $f \in [0,f_{lim}]$, being $f_{lim}=1MHz$, for the transfer impedance, both in modulus $|Z|(f,x_{concrete})$ [units of $1/h$] (c) and phase $\Phi_Z(f,x_{concrete})$ (d); in the hypothesis that $\Delta|Z|/|Z|=\Delta\Phi_Z/\Phi_Z=10^{-3}$, semi-logarithmic plots for both the inaccuracies (e) $\Delta\varepsilon_r/\varepsilon_r(f,x_{concrete})$, in the measurement of the permittivity $\varepsilon_r$, and $\Delta\sigma/\sigma(f,x_{concrete})$, of the conductivity $\sigma$, being the height/dimension ratio designed optimally in the Wenner linear ($x_{W,concrete}=0.087$) and square ($x_{S,concrete}=0.078$) configurations.

Fig. 9. With reference to a quadrupole probe designed by an electrode-electrode distance $L_0=1m$ and in a galvanic contact on a concrete of low electrical conductivity, i.e. $\sigma=10^{-4}$ S/m, $\varepsilon_r=4$; Bode's

diagrams, as function of the frequency $f$, for the transfer impedance, both in modulus $|Z|(f,L_0)$ (a) and phase $\Phi_Z(f,L_0)$ (b); semi-logarithmic plots for both the inaccuracies (c) $\Delta\sigma/\sigma(f)$, in the measurement of the conductivity $\sigma$, and $\Delta\varepsilon_r/\varepsilon_r(f)$, of the permittivity $\varepsilon_r$ $[\Delta|Z|/|Z|=\Delta\Phi_Z/\Phi_Z=10^{-3}]$.

Fig. 10. In the hypothesis that $\Delta|Z|/|Z|=\Delta\Phi_Z/\Phi_Z=10^{-3}$, referring to both the inaccuracies $\Delta\sigma/\sigma(\sigma,\varepsilon_r)$, for the electrical conductivity $\sigma$, and $\Delta\varepsilon_r/\varepsilon_r(\sigma,\varepsilon_r)$, for the dielectric permittivity $\varepsilon_r$, as functions of $\sigma$ and $\varepsilon_r$, and when the quadrupole probe is designed in the Wenner linear configuration, working in a fixed band $B=100kHz$, with an height/dimension ratio $x_{W,concrete}=0.087$, which is optimal for a capacitive contact only with a concrete of permittivity $\varepsilon_r=4$: plots for the orthogonal projections over the $(\sigma,\varepsilon_r)$ plane satisfying the conditions $\Delta\sigma/\sigma(\sigma,\varepsilon_r)\leq0.1$ (a) and $\Delta\varepsilon_r/\varepsilon_r(\sigma,\varepsilon_r)\leq0.1$ (b) [Tabs. 1, 2].

Fig. 11. With reference to a quadrupole probe, in galvanic contact, working in a fixed band $B=100kHz$, plots for the domains $(\sigma,\varepsilon_r)$ of the electrical conductivity $\sigma$ and the dielectric permittivity $\varepsilon_r$ such that: the transfer impedance is characterized by a modulus with a cut-off frequency $f_T=f_T(\sigma,\varepsilon_r)=\sigma/(2\pi\varepsilon_0(\varepsilon_r+1))$ ranging in the interval $f_T\in[100kHz,1MHz]$ (a); both the inaccuracies $\Delta\sigma/\sigma(\sigma,\varepsilon_r)$, in the measure of the conductivity $\sigma$, and $\Delta\varepsilon_r/\varepsilon_r(\sigma,\varepsilon_r)$, of the permittivity $\varepsilon_r$, result below a prefixed limit of $10\%$ $[\Delta|Z|/|Z|=\Delta\Phi_Z/\Phi_Z=10^{-3}]$ (b) [Tabs. 1, 3].